\documentstyle[11pt,aaspp4,psfig]{article} 

\def\deg{\ifmmode^\circ\else$^\circ$\fi}

\def\kpc{\ifmmode h^{-1}{\rm kpc}\else$h^{-1}{\rm kpc}$\fi}
\def\kms{\ifmmode {\rm km~s}^{-1}\else${\rm km~s}^{-1}$\fi}
\def\feh{\ifmmode {\rm [Fe/H]}\else [Fe/H]\fi}

\lefthead{Chiba et al.}
\righthead{Kinematics of Metal-Poor Stars}

\begin{document}

\title{Kinematics of Metal-Poor Stars in the Galaxy. III. Formation of the
Stellar Halo and Thick Disk as Revealed from a Large Sample of
Non-Kinematically Selected Stars}

\author{Masashi Chiba}
\affil{National Astronomical Observatory, Mitaka, Tokyo 181-8588, Japan\\
email: chibams@gala.mtk.nao.ac.jp}
\author{Timothy C. Beers}
\affil{Department of Physics \& Astronomy, Michigan State University, E.
Lansing, MI 48824\\email:  beers@pa.msu.edu}

\begin{abstract}

We present a detailed analysis of the space motions of 1203 solar-neighborhood
stars with metal abundances $\feh \le -0.6$, on the basis of a recently revised
and supplemented catalog of metal-poor stars selected without kinematic bias
(Beers et al. 2000). This sample, having available proper motions, radial
velocities, and distance estimates for stars with a wide range of metal
abundances, is by far the largest such catalog to be assembled to date.

We show that the stars in our sample with [Fe/H] $\le -2.2$,
which likely represent a ``pure'' halo component, are characterized by a
radially elongated velocity ellipsoid $(\sigma_U,\sigma_V,\sigma_W)=(141 \pm
11,106 \pm 9,94 \pm 8)$ km~s$^{-1}$ and small prograde rotation $<V_\phi>=30$
to 50 km~s$^{-1}$, consistent with previous analysis of this sample based on
radial velocity information alone (Beers \& Sommer-Larsen 1995).  In contrast
to the previous analysis, we find a decrease in $<V_\phi>$ with increasing
distance from the Galactic plane for stars which are likely to be members of
the halo population ($\Delta<V_\phi>/\Delta|Z|=-52 \pm 6$ km~s$^{-1}$
kpc$^{-1}$), which may represent the signature of a dissipatively formed
flattened inner halo.

Unlike essentially all previous kinematically selected catalogs, the metal-poor
stars in our sample exhibit a diverse distribution of orbital eccentricities,
$e$, with no apparent correlation between [Fe/H] and $e$.  This demonstrates,
clearly and convincingly, that the evidence offered by Eggen, Lynden-Bell, and
Sandage (1962) for a rapid collapse of the Galaxy, an apparent correlation
between the orbital eccentricity of halo stars with metallicity, is basically
the result of their proper-motion selection bias.  However, even in our
non-kinematically selected sample, we have identified a small concentration of
high-$e$ stars at [Fe/H] $\sim-1.7$, which may originate, in part, from
infalling gas during the early formation of the Galaxy.

We find no evidence for an additional thick disk component for stellar
abundances [Fe/H] $\le-2.2$. The kinematics of the intermediate-abundance stars
close to the Galactic plane are, in part, affected by the presence of a rapidly
rotating thick disk component with $<V_\phi> \simeq 200 $ km~s$^{-1}$ (with a
vertical velocity gradient on the order of $\Delta<V_\phi>/\Delta|Z|=-30 \pm 3$
km~s$^{-1}$ kpc$^{-1}$), and velocity ellipsoid
$(\sigma_U,\sigma_V,\sigma_W)=(46 \pm 4,50 \pm 4,35 \pm 3)$ km~s$^{-1}$.  The
fraction of low-metallicity stars in the solar neighborhood which are members
of the thick disk population is estimated as $\sim 10\%$ for
$-2.2 < $ [Fe/H] $ \le-1.7$ and $\sim 30\%$ for $-1.7< $ [Fe/H] $\le-1$.  We obtain an
estimate of the radial scale length of the metal-weak thick disk of $4.5 \pm
0.6$ kpc.

We also analyze the global kinematics of the stars constituting the halo
component of the Galaxy.  The outer part of the halo, which we take to be
represented by local stars on orbits reaching more than 5 kpc from the Galactic
plane, exhibits no systematic rotation.  In particular, we show that previous
suggestions of the presence of a ``counter-rotating high halo'' are {\it not}
supported by our analysis.  The density distribution of the outer halo is
nearly spherical, and exhibits a power-law profile that is accurately described
as $\rho \propto R^{-3.55 \pm 0.13}$.  The inner part of the halo is
characterized by a prograde rotation and a highly flattened density
distribution.  We find no distinct boundary between the inner and outer halo.

We confirm the clumping in angular-momentum phase space of a small number of
local metal-poor stars noted by Helmi et al. (1999). We also identify an
additional elongated feature in angular-momentum phase space extending from the
clump to regions with high azimuthal rotation.  The number of members in the
detected clump is not significantly increased from that reported by Helmi et
al., even though the total number of the sample stars we consider is almost
triple that of the previous investigation.  We conclude that the fraction of
halo stars that may have arisen from the precursor object of this clump may be
smaller than 10\% of the present Galactic halo, as previously suggested.

The implications of our results for the formation of the Galaxy are discussed,
in particular in the context of the currently favored Cold Dark Matter theory
of hierarchical galaxy formation.

\end{abstract}


\section{Introduction}

Over the past few decades, studies of the luminous halo population of
metal-deficient field stars and globular clusters have provided an increasingly
detailed picture of the formation and evolution of the Galaxy.  Because the
time required for mixing of the initial phase-space distribution of these
objects, via exchange of energies and angular momenta, is thought to exceed the
age of the Galaxy, kinematic information obtained at the present enables one to
elucidate the initial dynamical conditions under which these objects were born.
To the extent one is able to estimate ages of these halo population objects,
either directly (which is difficult at present), or indirectly (by postulating
that the {\it ensemble} metallicities of these objects increases with time),
their formation history is obtainable as well.  Thus, the dynamical and
chemical state of these halo-population objects provides important information
on how the Galaxy has developed its characteristic structures during the course
of its evolution.

Almost forty years ago, the ``canonical'' scenario of the early dynamical
evolution of the Galaxy was put forward by Eggen, Lynden-Bell, \& Sandage
(1962, hereafter ELS) to explain what they believed to be an observed
correlation between the orbital characteristics and metal abundances of stars
in the solar neighborhood.  Focusing on the lack of metal-poor stars with low
eccentricity orbits in their (proper-motion selected) sample, ELS argued that
the Galaxy must have undergone a rapid collapse, then formed a rotationally
supported disk.  Although  criticism of the ELS model has been levied because
of the potential influence of their kinematic selection bias, especially the
extent to which this might alter the derived collapse timescale of the early
Galaxy (Yoshii \& Saio 1979; Norris, Bessell, \& Pickles 1985; Norris 1986;
Norris \& Ryan 1991; Beers \& Sommer-Larsen 1995, hereafter BSL; Chiba \&
Yoshii 1998, hereafter CY), the ELS collapse picture has long been influential
for studies of disk galaxies like our own, and for elliptical galaxies as well
(e.g., Larson 1974; van Albada 1982).

An alternative picture for the formation of the Galactic halo was proposed by
Searle \& Zinn (1978, hereafter SZ), who noted a number of difficulties in
reconciling several observed properties of the halo globular cluster system
with predictions of the ELS model. Among these, the existence of a large
(several Gyr) spread in the inferred ages of the Galactic globulars, and the
lack of an abundance gradient with distance from the Galactic center were
thought to be the most crucial.  SZ suggested that, in its earliest epochs, the
halo component of the Galaxy may have experienced prolonged, chaotic accretion
of subgalactic fragments.  More recent studies of halo field stars also provide
evidence which may support the SZ picture, including a possible age spread
among halo subdwarfs (e.g., Schuster \& Nissen 1989; Carney et al. 1996), a
gradient in the inferred ages of field horizontal-branch (FHB) stars and
RR Lyrae variables, in the sense that the outer halo FHB stars and RR Lyrae
variables appear several Gyr younger than those of the inner halo (Preston,
Shectman, \& Beers 1991; Lee \& Carney 1999), a report of the apparent
clustering of FHB stars in the halo (Doinidis \& Beers 1989), possible
kinematic substructure in the halo (e.g., Doinidis \& Beers 1989; Majewski,
Munn, \& Hawley 1994; 1996), and distinct changes in the kinematics of the
field populations as one moves from the inner to outer halo (e.g., Majewski
1992; Carney et al. 1996; Sommer-Larsen et al. 1997).

In order to assess which picture, ELS or SZ (or both, e.g., Norris 1994;
Freeman 1996; Carney et al. 1996), more correctly describes the early
history of the Galaxy, we require a large and reliable set of data for
halo-population objects chosen with criteria that do not unduly influence the
subsequent analysis.  As we show, the analysis of stars chosen with a kinematic
selection bias can be particularly troublesome.  Interestingly, in order to
obtain adequately large samples of stars exhibiting a range of metallicities in
the solar neighborhood, an abundance bias is actually {\it required}, otherwise
the exceedingly rare very low-metallicity stars of the halo population will not
be represented in sufficient numbers.  One must exercise caution, however, that
abundance estimates for stars in the sample under consideration are as accurate
as possible, due to the presence of an overlap of the local halo with the
relatively high density thick-disk population  (e.g., Anthony-Twarog \& Twarog
1994; BSL; Ryan \& Lambert 1995; Twarog \& Anthony-Twarog 1996; CY). It is
similarly important to assemble a large and homogeneously analyzed sample, both
to minimize statistical fluctuations in the derived kinematic quantities, and
to reduce the effects of other systematic errors (such as might arise in
estimates of stellar distances).

In this paper we present an analysis of the kinematics of metal-deficient
field stars in the solar neighborhood, based on a large catalog of stars
selected without kinematic bias (Beers et al. 2000, hereafter Paper II).  This
catalog, consisting of 2041 stars from the published literature with
abundances [Fe/H]$ \le -0.6$, includes updated stellar positions, newly derived
homogeneous distance estimates, revised radial velocities, and refined metal
abundance estimates.  Moreover, a subset of some 1200 stars in the catalog now
have available proper motions, taken from a variety of recently completed
proper motion catalogs. We note that this catalog is (by far) the largest
sample of metal-deficient field stars with available proper motions among any
previously assembled non-kinematically selected samples.  Thus, it is now
possible to draw a much more definitive picture of the early kinematic
evolution of the Galaxy.

Our paper is organized as follows. In \S 2 we present a discussion of the
detailed velocity distributions of our sample stars, concentrating on those
presently located in the solar neighborhood.  In \S 3 we analyze the orbital
motions of these stars.  In \S 4 we consider the global character of the halo
of the Galaxy, as deduced from the kinematics of a local sample.  In \S 5, we
further examine evidence for kinematic substructure in the phase-space
distribution of the halo. Finally, in \S 6, the results of the present work are
summarized, and their implications for the formation and evolution of the
Galaxy are discussed.

\section{Velocity Distributions of the Metal-Poor Stars}

\subsection{Individual and Mean Space Velocities}

Paper II of this series presented proper motions for 1214 stars with
[Fe/H] $ \le-0.6$, as well as for a number of slightly more metal-rich stars.
Within this sample, 1203 stars with [Fe/H] $\le-0.6$ have distance estimates and
radial velocities as well as proper motions, so that the full three-dimensional
velocities are directly calculable.  Figure 1 is a reproduction of Figure 8
from Paper II, and shows the local velocity components vs. [Fe/H] for these
1203 stars.  The velocity components $U$, $V$, and $W$ are directed to the
Galactic anticenter, rotation direction, and north Galactic pole, respectively,
and have been corrected for the local solar motion $(U_\odot, V_\odot,
W_\odot)=(-9,12,7)$ km~s$^{-1}$ with respect to the local standard of rest
(LSR) (Mihalas \& Binney 1981).  Note the excellent coverage of this sample
over the entire range of Galactic abundances, especially below [Fe/H] $=-2$ and
above [Fe/H] $=-1$, where previous studies that made use of, for example, the
sample of metal-poor stars studied with {\it Hipparcos}, lack sufficient
numbers of stars (compare, e.g., with Figure 4 of CY).

To examine the characteristic {\it local} velocity distributions of our sample,
we confine ourselves to a discussion of the stars for which the Galactocentric
distance along the plane, $R$, is between 7 and 10 kpc, and those for which the
distance from the Sun, $D$, is within 4 kpc.  Six stars in this subsample have
large rest-frame velocities, $V_{RF}>550$ km s$^{-1}$, that are in excess
of the canonical escape velocity in the solar neighborhood ($V_{esc} \sim 500-
550$ km s$^{-1}$; Carney, Latham, \& Laird 1988).  Although some of the space
velocities may indeed be this high, the majority of these stars probably have
large $V_{RF}$ due to an over-estimation of their distances, which has
artificially inflated their estimated tangential velocities.  We choose to
remove these extreme-velocity stars by placing an additional limit of
$V_{RF}\le 550$ km s$^{-1}$ on the sample.  The stars satisfying the above
selection criteria are referred to as the ``Selected Sample'' in the following
discussion.

We first calculate the mean velocities $(<U>,<V>,<W>)$ and velocity dispersions
$(\sigma_U,\sigma_V,\sigma_W)$ for the Selected Sample -- values for five
characteristic ranges in metal abundance are listed in Table 1a.  Velocity
dispersions are estimated from the standard deviations, after correction for
the typical measurement errors in the velocities ($\sim 10$ km~s$^{-1}$).
Figure 2 shows $(\sigma_U,\sigma_V,\sigma_W)$ as a function of [Fe/H].  In this
Figure we have adopted a finer binning in metal abundance; the dispersion
measurements in each bin are listed in Table 1b.  The filled and open circles
in Figure 2 denote the stars at $|Z|<1$ kpc and $|Z|<4$ kpc, respectively,
where $Z$ is the height above the Galactic plane.

The most metal-deficient stars in the Selected Sample, those more metal-poor
than [Fe/H] $=-2.2$, are dominated by members of the halo population.  For $|Z|
< 1$ kpc, these stars exhibit a radially elongated velocity ellipsoid
$(\sigma_U,\sigma_V,\sigma_W)= (141\pm11,106\pm9,94\pm8)$ km~s$^{-1}$, in good
agreement with previous results (e.g., BSL; CY).  With a slightly more
metal-rich cut on the abundances, i.e., selecting stars with [Fe/H] $ <-1.7$, we
obtain similar values, $(148\pm7,110\pm5,92\pm4)$ km~s$^{-1}$, so it appears
that the shape of the velocity ellipsoid remains essentially unchanged with
varying [Fe/H] below [Fe/H] $=-1.7$. In this regard Norris (1994) claimed,
from his analysis of a sample of high proper-motion stars, that $\sigma_W$
continues to increase with decreasing [Fe/H], even at its lowest levels.  This
result was taken to indicate the possible existence of a dynamically ``hot''
proto-disk population at low abundances.  Carney et al.  (1996) disputed this
result, as their analysis of a different set of high proper-motion stars
indicated that the disk component is not dynamically hot, at least when
membership is confined to the stars orbiting exclusively within the inner part
of the Galaxy, $R \le 14$ kpc.  Figure 2c shows no evidence for an increase of
$\sigma_W$ at low abundances.

The velocity dispersion components of the Selected Sample in the more
metal-rich abundance ranges decrease as the contribution of the thick disk
component progressively increases. In particular, for $-0.7\le$ [Fe/H] $<-0.6$
and $|Z|<1$ kpc, where the contribution of the halo component is expected to be
negligible, the mean $V$ velocity, $<V>$, is $-20 \pm 5$ km~s$^{-1}$; the
velocity dispersions are $(\sigma_U,\sigma_V,\sigma_W)=(46\pm4,50\pm4,35\pm3)$
km~s$^{-1}$ . This result is in agreement with previously derived kinematic
parameters for the thick disk, which appears to be in rapid rotation ($\sim$200
km~s$^{-1}$), provided the rotational speed of the LSR is $V_{LSR}=220$
km~s$^{-1}$ (BSL).  With these values for the thick disk kinematics, it is
possible to estimate the radial scale length of this component using the
following formula (Binney \& Tremaine 1987):

\begin{equation}
2 V_{LSR} V_{lag} -V_{lag}^2 = \sigma_U^2
 \left( -1 + \frac{\sigma_V^2}{\sigma_U^2} + 2 \frac{R}{h_R} \right) \ ,
\end{equation}

\noindent where $V_{lag}$ is the asymmetric drift given by $V_{lag}=-<V>$, and
$h_R$ is the scale length of the disk, provided its density varies as
$\exp(-R/h_R)$.  By inserting $V_{LSR}=220\pm10$ km~s$^{-1}$, the assumed
solar radius $R=8.5$ kpc (Kerr \& Lynden-Bell 1986)
and derived parameters for $-0.7\le$ [Fe/H] $<-0.6$ in Equation (1), we
obtain $h_R=4.5 \pm 0.6$ kpc. This is in good agreement with $h_R=4.7\pm0.5$
kpc obtained by BSL, and also with the lower limit of $h_R\simeq4.5$ kpc
derived by Ratnatunga \& Freeman (1989). We note that the second term on the
left-hand side of Equation (1) has been omitted in some previous works, as
it is small compared to other terms. If we were to exclude this term, we would
obtain $h_R=4.3\pm0.6$ kpc, thereby slightly underestimating $h_R$.

\subsection{Rotational Character of the Selected Sample}

We now examine the rotational character of the Selected Sample.  Figure 3 shows
the mean rotational velocities $<V_\phi>$ (the rotation velocity in a
cylindrical coordinate frame) as a function of [Fe/H], based on stars in the
Selected Sample.  The left-hand panel in Figure 3a displays the results for the
abundance ranges listed in Table 2, for three subsets of the sample as a
function of distance above the plane.  In the right-hand panel of Figure 3a,
the bins are obtained by passing a box of width $N=100$ stars,
ordered by metallicity, with an overlap of 20 stars each. The latter approach
is adopted to avoid any effects of the arbitrary placement of bins on the
results.

Note that the panels in Figure 3a are obtained with full knowledge of the space
motions of the stars in the Selected Sample.  Figure 3b is based on the radial
velocities alone, applying the methodology of Frenk \& White (1980) (FW) (see
also Norris 1986; Morrison, Flynn, \& Freeman 1990, hereafter MFF; BSL). The
solid lines in the left and right-hand panels of Figure 3b denote $<V_\phi>$ as
derived for the stars with available proper motions (denoted as
$<V_\phi>_{pm}^{FW}$ in Table 2), i.e., the same sample as in Figure 3a,
whereas the dashed lines are for all of the stars with available radial
velocities ($<V_\phi>_{all}^{FW}$ in Table 2).  Comparison between Figures 3a
and 3b allows us to examine whether the Selected Sample is subject to any
significant kinematic bias, since, if so, the $<V_\phi>$ derived from the space
motions is expected to be systematically smaller from that determined on the
basis of radial velocities alone (Ryan \& Norris 1991; Norris \& Ryan 1991).

Figure 3a clearly indicates that the rotational properties of the Selected
Sample change discontinuously at [Fe/H] $\simeq -1.7$.  Stars with [Fe/H]
$<-1.7$ exhibit no systematic variation of $<V_\phi>$ with decreasing [Fe/H].
It is interesting to note that the subsample of low-abundance stars with
$|Z|<1$ kpc show a rather large prograde rotation of $<V_\phi>=30\sim50$
km~s$^{-1}$, and that $<V_\phi>$ decreases if stars at larger heights are
considered, at least for the two abundance bins centered on [Fe/H] $ = -1.9$ and
$-2.2$, respectively.  This behavior is not exhibited, however, for stars in
the lowest abundance bin.  To check the significance of this feature, we have
combined the stars in our Selected Sample in the metallicity interval
$-2.4\le{\rm [Fe/H]} \le -1.9$ and obtained $<V_\phi>$ by sweeping a box of 50
stars ordered by $|Z|$, with an overlap of 30 stars. The results are summarized
in Table 3 and depicted in Figure 4. The lower solid line in Figure 4 is a
least-squares fit to the data, which yields $\Delta<V_\phi>/\Delta|Z|=-52 \pm
6$ km~s$^{-1}$ kpc$^{-1}$, indicating the presence of a significant
vertical gradient in $<V_\phi>$ at low abundances.  Figure 4 also suggests that
$<V_\phi>$ beyond $|Z|\sim1.2$ kpc has a nearly constant zero value.  If we
exclude the last point at $|Z|=1.76$ kpc from the fit, we obtain
$\Delta<V_\phi>/\Delta|Z|=-62 \pm 5$ km~s$^{-1}$ kpc$^{-1}$ (shown as a dotted
line).  Note that Majewski (1992) reported evidence for a halo component which
is in retrograde motion ($V_{\phi} = -275 \pm 16 $ km~s$^{-1}$), but which
exhibited no gradient of rotation with distance from the plane, a result that
is clearly at odds with our present result.

Figure 3a shows that for stars in the Selected Sample with [Fe/H] $>-1.7$ there
is a clear linear dependence of $<V_\phi>$ on [Fe/H], a dependence that remains
essentially unchanged even if the range of $|Z|$ is varied.  This discontinuity
has been seen before, of course, based on analysis of smaller samples (e.g.,
Norris 1986; Carney 1988; Zinn 1988; Norris \& Ryan 1989; BSL; CY).  The
inescapable conclusion is that the transition from halo to disk must not have
occurred in a continuous manner, as predicted in the ELS model.  However, the
vertical gradient in $<V_\phi>$ for metal-poor stars noted above suggests that
the halo was not formed in a totally chaotic, dissipationless manner as implied
in the SZ hypothesis.  Rather, dissipational processes may have played a role
in the initial contraction of the halo, likely involving energy exchange with
the gas phase (Carney et al. 1996).

It is also worth noting from Figure 3 that the stars of the thick disk, which
dominate the Selected Sample for [Fe/H] $>-1$, exhibit only a small change in
$<V_\phi>$ with increasing $|Z|$.  Table 3 and Figure 4 summarize the change of
$<V_\phi>$ with $|Z|$ for stars likely to be dominated by the thick disk, i.e.,
in the abundance range $-0.8\le$ [Fe/H] $\le-0.6$.  The least-squares fit to the
data (as shown by upper solid line) yields $\Delta<V_\phi>/\Delta|Z|=-30 \pm 3$
km~s$^{-1}$ kpc$^{-1}$, much smaller than the gradient obtained for halo stars
with $-2.4\le$ [Fe/H] $\le-1.9$, but similar to previous estimates of the
thick disk rotational velocity gradients reported by Majewski (1992).

The relation between $<V_\phi>$ and [Fe/H] as shown in the panels of Figure 3a
is also seen in the panels of Figure 3b. In particular, $<V_\phi>$ obtained
from the subsample having available proper motions (dashed lines) is
essentially the same as that from the entire sample (solid lines) within
standard errors in $<V_\phi>$ (except for [Fe/H] $\sim-1.6$: see below).
This is consistent with our argument given in Paper II that the subsample
based on stars with available proper motions is not subject to any significant
kinematic bias.

We note that near [Fe/H] $=-1.6$ in Figure 3b, $<V_\phi>$ obtained from
consideration of the sample with full space motions is {\it larger} than that
obtained from the sample using radial velocities alone. The apparent
retrograde rotation of globular clusters in the similar metallicity range
was also reported by Rodgers \& Paltoglou (1984) based on radial velocities
alone, whereas Dinescu, Girard, \& van Altena (1999) found no sign of
significant retrograde rotation using full space motions of globular clusters.
It is worth noting that this difference in the rotational
velocity using either of radial velocities or full space motions is not
a signature of kinematic bias, since the result is in the opposite direction to
that expected.  This difference probably arises due to limitations of the FW
methodology as applied to our sample.  The FW method implicitly assumes that
the angle, $\psi$, between the line-of-sight and the rotational direction is
randomly distributed in the sample, which may not apply in this case.  In
addition, the apparent excursion to large retrograde rotation,
$<V_\phi>\simeq-40$ km~s$^{-1}$ near [Fe/H] $=-1.6$ in the right-hand panel of
Figure 3b (also noted by BSL), may be caused by a few outliers having large
velocities as seen by an observer at rest with respect to the Galactic center,
$V_{gal}$.  If we exclude five stars having large $V_{gal}$ from the bin
centered at [Fe/H] $=-1.6$,  we obtain $<V_\phi>=-12$ km~s$^{-1}$.  We have
verified that the effect of outliers on $<V_\phi>$ is small in other abundance
ranges.

The influence of a disk-like population for [Fe/H] $ > -1.7$ is certainly
suggested by the appearance of Figure 3.  The question of the limiting
abundance of a so-called metal-weak thick disk (MWTD)\footnote{These
metal-poor stars with disk-like kinematics may also include a considerable
portion of the thin disk if its metallicity distribution overlaps that of the
thick disk (Wyse \& Gilmore 1995).} has been considered
several times in the past (MFF; Rodgers \& Roberts 1993; Layden 1995; BSL; Ryan
\& Lambert 1995; Twarog \& Anthony-Twarog 1996; CY).  Figure 5 shows the
frequency distribution of $V_\phi$ for the stars in the Selected Sample with
available space motions, for subsets chosen to have (a) $|Z|<1$ kpc, and (b)
$|Z|\ge 1$ kpc, respectively.  At $|Z|<1$ kpc, where the disk-like kinematics
are expected to be more evident than at larger heights above the plane, the
metal-rich stars with [Fe/H] $>-1$ are peaked at $V_\phi=200$ km~s$^{-1}$, a
rather high rotational velocity.  One also sees the presence of a small
contribution of the stars with halo-like kinematics. At lower abundances the
halo-like kinematics become much more dominant; the contribution of the MWTD is
apparently decreasing at lower abundances and higher $|Z|$.

To quantify the fraction of the MWTD in our local sample within the specified
abundance ranges, we have fit the subset of stars with $|Z|<1$ kpc using a
mixture of two Gaussian distributions for $V_\phi$, representing the halo and
disk populations. The halo kinematic parameters ($<V_\phi>_{halo},
\sigma_{\phi,halo}$) = (+33,106) km~s$^{-1}$ are derived from stars with
[Fe/H] $\le-2.2$. The rotation velocity of the disk component, $<V_\phi>_{disk}
= +200$ km~s$^{-1}$, is obtained considering the stars in the metallicity range
$-0.7< $ [Fe/H] $\le-0.6$.  With these parameters fixed, we evaluate the most
likely values of the velocity dispersion of the disk, $\sigma_{\phi,disk}$ and
fraction $F$ of the MWTD, using a maximum likelihood analysis (see also MFF;
CY).  The likelihood function for the stars with $V_\phi^i$ is given by

\begin{equation}
\log f(F,\sigma_{\phi,disk}) = \sum_{i=1}^N \log[ F f_{disk}^i
 + (1-F) f_{halo}^i] \ ,
\end{equation}

\noindent where $f_{disk}^i$ ($f_{halo}^i$) denote Gaussian functions with
mean velocities $<V_\phi>_{disk}$ ($<V_\phi>_{halo}$) and dispersions
$\sigma_{\phi,disk}$ ($\sigma_{\phi,halo}$).  The results of the likelihood
analysis are tabulated in Table 4, and  shown by the solid curves in Figure
5.  The MWTD contributes about 30\% of the metal-poor stars in the
abundance range $-1.7<$ [Fe/H] $\le-1$, which is smaller than the fraction
derived by MFF ($\sim$72\%) and BSL ($\sim$60\%), but larger than the result of
CY ($\sim$10\%). The fraction of the MWTD is quite modest in the more
metal-poor ranges, in contrast to the suggestion of BSL, who argued for
$\sim$30\% even at [Fe/H] $<-2$. One reason that our result may differ so
strikingly from that of BSL is that the estimated abundances for many of the HK
survey stars at [Fe/H] $<-2$, listed in the original BSL catalog, are
likely to have been underestimated by $\sim 0.3$ dex (see Figure 1 of Paper
II).

The RR Lyrae stars in our Selected Sample (shown as shaded histograms in Figure
5) exhibit no clear disk-like kinematics, even in the intermediate abundance
range of $-1.7<$ [Fe/H] $\le-1$.  This confirms earlier results by MFF, Layden
(1995), and CY, and may imply a somewhat younger age of the thick disk as
compared to the halo.  However, the numbers of RR Lyraes with available space
motions is rather small, so this question, especially in conjunction with their
period distributions to investigate different populations of RR Lyraes
(Lee \& Carney 1999), should be revisited when the sample size has been
increased.

\section{Orbital Properties of the Metal-Poor Stars}

In this section we investigate the orbital properties of our sample of
stars in a given Galactic potential.  We adopt the analytic St\"ackel-type
potential developed by Sommer-Larsen \& Zhen (1990, hereafter SLZ), which
consists of a flattened, oblate disk and a nearly spherical massive halo. This
model potential is consistent with the mass model of Bahcall, Schmidt, \&
Soneira (1982), exhibiting a flat rotation curve beyond $R=4$ kpc, and having a
commensurate local mass density at $R = R_\odot$.  In contrast to a
non-analytic potential, for which numerical integrations of orbits are
required, the analytic nature of the adopted potential has the great advantage
of maintaining clarity in the analysis, as demonstrated below. In the Appendix,
we summarize the properties of the St\"ackel mass model, and provide
expressions for three integrals of motion in such a model (see also de Zeeuw
1985; Dejonghe \& de Zeeuw 1988)

\subsection{The Relationship between Orbital Eccentricity and Metal Abundance}

We first compute orbital eccentricities, defined as $e = (r_{ap}-r_{pr}) /
(r_{ap}+r_{pr})$, where $r_{ap}$ and $r_{pr}$ denote the apogalactic and
perigalactic distances of the orbits, respectively.  These orbital parameters
are tabulated in Table 3 of Paper II for the stars under consideration. In
Figure 6a, we show the relation between $e$ and [Fe/H]. As is evident, there is
{\it no strong correlation} between these quantities, and the metal-poor stars
below [Fe/H] $=-2$ exhibit a diverse range in orbital eccentricities.  This is
in sharp contrast to the ELS result, and confirms previous suggestions from a
number of workers, but in a much more definitive manner (Yoshii \& Saio 1979;
NBP; Carney \& Latham 1986; Carney, Latham, \& Laird 1990; Norris \& Ryan 1991;
CY).  In addition to the diverse distribution of $e$ at all abundances, we note
a small concentration of the stars at $e \sim 0.9$ and [Fe/H] $\sim-1.7$, which
is somewhat reminiscent of the original ELS result.  It is perhaps not
coincidental that the excess number of high-$e$ stars occurs at an abundance
which matches the sharp discontinuity of $<V_\phi>$ found at [Fe/H] $=-1.7$,
where $<V_\phi>$ is almost zero (Figure 3). This may suggest that a significant
fraction of the metal-poor stars with abundances near [Fe/H] $ = -1.7$ formed
from infalling gas of this metallicity during an early stage of Galaxy
formation, in a manner similar to an ELS collapse.

In Figure 6b, we show the mean eccentricity, $<e>$, vs. [Fe/H], where the bins
are obtained by passing a box of width $N=100$ stars, ordered by metallicity,
with an overlap of 20 stars each.  For comparison, the dashed line denotes the
result of Carney et al. (1996) for their high proper motion sample.  There is a
clear difference from our results -- the use of a kinematically selected sample
{\it overestimates} the average orbital eccentricities at a given [Fe/H], by an
amount up to 0.2.

Figure 7 shows the cumulative distributions of $e$, $N(<e)$, in two specific
abundance ranges, (a) for [Fe/H] $\le-2.2$, and (b) for $-1.4<$ [Fe/H] $\le-1$.
Figure 7a clearly demonstrates that even at quite low abundance, roughly 20\%
of our stars have $e<0.4$.  The different lines correspond to the cases when
the range of $|Z|$ is changed.  It is apparent that, for stars with
[Fe/H] $\le-2.2$, the cumulative distribution function of $e$ is unchanged when
considering subsets of the data with a range of $|Z|$, suggesting the absence
of any substantial disk-like component below this metallicity.  By way of
contrast, Figure 7b shows that stars with intermediate abundances exhibit (a) a
higher fraction of orbits with $e<0.4$ than for the lower abundance stars, (b)
a decrease in the relative fraction of low eccentricity stars as larger heights
above the Galactic plane are considered, and (c) convergence at larger heights
to a fraction which is close to the 20\% obtained for the lower abundance
stars.  These results imply that the orbital motions of the stars in the
intermediate abundance range are, in part, affected by the presence of
thick-disk component with a finite scale height.  We recall that CY and Chiba,
Yoshii \& Beers (1999), using a sample of metal-poor stars with {\it Hipparcos}
measurements, found a further decrease of the fraction of the stars with
$e<0.4$ at larger $|Z|$, without achieving the convergence noted here (see
Figure 15 of CY).  This was presumably due to the lack of a sufficient number
of intermediate abundance stars at large $|Z|$ in the sample considered by
these authors.

\subsection{Structural Parameters of the MWTD Component}

We now seek to quantitatively describe the abundance range, scale height, and
fraction of the MWTD component in our sample. We apply a Kolmogorov-Smirnoff
(KS) test of the null hypothesis that the differential distributions of $e$,
$n(e)$, for stars in a specified abundance range, are drawn from the same
parent population of eccentricities as stars belonging to a ``pure'' halo
component.  Based on our analysis above, we take the subsample of 78
stars with [Fe/H] $\le-2.2$ and $|Z|<1$ kpc to represent the pure halo
component.  We then calculate the KS probabilities, $P_{KS}$, for the stars in
various intermediate abundance ranges and with $|Z|>Z_{lim}$, where $Z_{lim}$
is the lower limit on the heights of the stars above the Galactic plane.  We
expect that, even if the specified abundance range is contaminated by stars
with disk-like kinematics, it will be dominated by halo-like kinematics above
{\it some} $|Z|=Z_{lim}$, with $P_{KS}$ exceeding 0.2 (i.e., the subsamples
being consistent with draws from the same parent population of orbital
eccentricities).

Figure 8 shows the results of the KS tests.  In order to obtain an estimate for
the value of $Z_{lim}$ above which the populations cannot be distinguished,
Figure 8a depicts the results for the stars below [Fe/H] $=-1$, but above the
specified {\it lower limit} for the abundance.  In the left-hand panel of
Figure 8a, it is seen that $P_{KS}$ rapidly increases at $Z_{lim}=0.5$ to 1.3
kpc and then remains roughly constant at larger $Z_{lim}$.  In the right-hand
panel of Figure 8a, the distribution of $P_{KS}$ on $Z_{lim}$ changes
dramatically as one passes from the inclusion of stars with [Fe/H] $ > -1.9$ to
those with [Fe/H] $ > -2.0$.  When stars with abundances as low as [Fe/H] $ =
-2.0$ are included, there is {\it no value} of $Z_{lim}$ for which the
distributions can be distinguished.  To identify the lower limit on the
abundance of stars which are members of the MWTD, Figure 8b shows the
distribution of $P_{KS}$ for metal-poor stars above [Fe/H] $=-2.2$, but below
the specified {\it upper limit} on abundance.  As is seen in the left-hand
panel of this figure, $P_{KS}$ remains small (indicating that the populations
can be distinguished) at all $Z_{lim}$.  In the right-hand panel of Figure 8b,
one sees that although there exists a region at small $Z_{lim}$ where the
populations can be marginally distinguished when the upper limit on abundance
is taken to be [Fe/H] $= -1.9$, there is no such region when stars with an
upper limit of [Fe/H] $= -2.0$ is considered.  These results suggest that the
MWTD component has a characteristic scale height of roughly 1 kpc, above which
halo-like orbital motions dominate, and a lower abundance limit near
[Fe/H] $=-2.0$. We note here that the number of stars employed in the KS tests
is sufficiently large in the ranges of $Z_{lim}$ considered (e.g. $N=240$ for
$-2<$ [Fe/H] $\le-1$ and $N=54$ for $-2.2<$ [Fe/H] $\le-2$ at $Z_{lim}=1$ kpc).
However, it would be useful, especially at larger $Z_{lim}$, to boost the
sample sizes so that more detailed investigations can be carried out.

We now estimate the contribution of the MWTD component in the solar
neighborhood, $F$, using the distribution of $e$ in various abundance ranges.
Following the methodology developed by CY, we perform a Monte Carlo simulation
to predict the $e$-distribution from a mixture of stars contributed by the
thick-disk and halo populations, adopting the kinematic parameters for these
components derived in
\S 2, and compare with the observed cumulative distribution functions of
eccentricity in our sample with $|Z| < 1$ kpc.  Figure 9 shows the results of
this exercise.  As is clear, more metal-rich ranges are described by a larger
$F$, but in the abundance range $-2.2<$ [Fe/H] $\le-2$, $F \simeq 0$, in good
agreement with the results obtained from comparison of the differential
eccentricity distributions.  Figure 10 shows the dependence of $F$ on [Fe/H]
for stars with $|Z|<1$ kpc, where the fits are made with bins of $0.2$ dex for
[Fe/H] and $0.1$ for $F$.  The $F$([Fe/H]) derived here, based on full
knowledge of the stellar orbital motions, is rather similar to that found by
BSL based on radial velocities alone, except for the abundance range below
[Fe/H] $<-1.5$, where the MWTD appears more modestly populated.  A
characteristic value of $F=0.3$ over the abundance range $-1.7<$ [Fe/H] $\le-1$
is obtained from our present analysis.

Having obtained the structural parameters of the MWTD component, we extract a
set of likely members of the MWTD component in our sample.   High-resolution
spectroscopic observations of these stars, to obtain estimates of their
individual elemental abundances, should provide valuable information concerning
the nature of MWTD stars, and reveal differences, if any, in their compositions
relative to similar metallicity stars of the halo population (e.g., Bonifacio,
Centurion, \& Molaro 1999).   In Table 5 we list the stars satisfying (1)
$-2.2\le$ [Fe/H] $\le-1$, (2) $|Z|\le 1$ kpc, (3) $V_\phi\ge<V_\phi>_{disk} -
\sigma_{\phi,disk} $, and (4) $|V_R|\le
\sigma_{R,disk}$ and $|V_Z|\le \sigma_{Z,disk}$, where $<V_\phi>_{disk}=200$
km~s$^{-1}$ and $(\sigma_{R,disk},\sigma_{\phi,disk},\sigma_{Z,disk})=
(46,50,35)$ km~s$^{-1}$, as derived above). Condition (3) corresponds to the
high rotation velocity of the candidate members and condition (4) is placed
so that their velocities in $R$ and $Z$ directions are confined within
a 1 $\sigma$ range relative to the zero mean, i.e. within the velocity
dispersions of the MWTD component.  In Table 5, the fourth column denotes the
classification of each stellar type. We follow the coding of paper II  -- D:
main-sequence dwarf star; TO: main-sequence turnoff star; SG: subgiant star; G:
giant star; AGB: asymptotic giant branch star; FHB: field horizontal-branch
star; RRV: RR Lyrae variable star; V: variable star.  Note that Table 5
supersedes Table 8 of BSL, as we now have much more complete kinematic
information.

\section{Global Dynamics and Structure of the Halo}

Although our present sample is dominated by stars located in the vicinity of
the Sun, the orbits of many of these stars explore regions well into the
more distant halo of the Galaxy.  Thus, their local kinematics provide
information on the global dynamics and structure of the halo (May and Binney
1986).  In this section we first investigate the rotational properties of the
halo at large heights from the Galactic plane, then use this same sample to
obtain a picture of the global density distribution of the halo.

\subsection{Rotational Properties of the Halo}

Majewski (1992) claimed, on the basis of measured proper motions for an {\it in
situ} sample of halo subdwarfs located at $Z > 5$ kpc (in a small field in the
direction of the North Galactic Pole), that stars at such large heights above
the Galactic plane exhibit a net {\it retrograde} rotation $<V_\phi>
\simeq -55 \pm 16$ km~s$^{-1}$, in contrast to the stars nearer the
plane, which show a near-zero or slightly prograde rotation.  Although there
have been criticisms of this result (e.g., Ryan 1992), a number of workers have
also reported observations of separate samples which seem to support this view,
so the true situation has remained unclear.  For example, Carney et al. (1996)
reported evidence for a retrograde rotation in the subset of their local sample
of high proper-motion stars whose orbits extend far above the plane.  In their
analysis, they divided the sample into those stars with $Z_{max}\le2$ kpc, and
$Z_{max}\ge5$ kpc, respectively, where $Z_{max}$ is the maximum distance of the
derived orbit from the plane.  Carney et al. showed that their ``high halo''
sample, with $<$ [Fe/H] $>=-2.04$ and $Z_{max}\ge5$ kpc, exhibited a net
retrograde rotation ($<V_\phi>=-45 \pm 22$ km~s$^{-1}$), whereas their ``low
halo'' sample at $Z_{max}\le2$ kpc exhibited a net prograde rotation ranging
from $<V_\phi>=12$ km s$^{-1}$ to 44 km~s$^{-1}$, depending on the specific
criteria chosen to avoid stars of the disk component.

Figure 11a reproduces the original data of Carney et al. (1996), for
stars with [Fe/H] $\le-1.5$. Since, in their estimates of $Z_{max}$, Carney et
al. adopted a different Galactic potential than ours, we have also
re-determined $Z_{max}$ for their sample with the same potential described in
\S 3, and show the results in this same figure.  Regardless of the adopted
potential, it is apparent that the stars at $Z_{max}\ge5$ kpc exhibit a net
retrograde rotation, as compared to the stars at $Z_{max}\le2$ kpc, which are
in prograde rotation.  Carney et al. argued that this result might be explained
by the presence of two distinct halo populations, a high halo formed via
accretion of fragments (such as in the SZ model), and a low halo formed
from an organized contraction (similar to the ELS model).

The Carney et al. sample is based on high proper-motion stars selected from the
Lowell Proper Motion Catalog, where proper motions are measured to exceed
0.26'' yr$^{-1}$, and the New Luyten Two-Tenths Catalog with proper motions
exceeding roughly 0.18'' yr$^{-1}$. Thus, their sample is {\it a-priori} biased
against inclusion of stars with prograde rotation close to the velocity of the
LSR.  What remains to be evaluated is the effect that this kinematic bias may
have on the observed kinematics of the high halo.

We consider the question of bias in the Carney et al. sample via a Monte Carlo
simulation, with the following assumptions.  ``Stars'' with [Fe/H] $ < -1.5$
are randomly distributed in Galactic coordinates $(l,b)$, and are assumed to
have a Gaussian velocity distribution with
$(\sigma_U,\sigma_V,\sigma_W)=(141,108,94)$ km~s$^{-1}$, but with no systematic
rotation.  Distances to the simulated stars are taken by adopting a Gaussian
form with mean 0.18 kpc and dispersion 0.09 kpc, which reproduces well the
distance distribution of their sample stars.  We further assume that only the
stars whose inferred proper motions exceed 0.26'' yr$^{-1}$ are observed at the
Sun with $V_{LSR}=220$ km s$^{-1}$.  The result is shown in panel Figure 11b.
The simulated data sample exhibits a net retrograde rotation for stars in the
high halo.  We note that our simulation does {\it not} exhibit a net prograde
rotation in the low halo, as seen in the Carney et al. sample.  This may arise
because their sample is not distributed randomly in $(l,b)$, as our simulation
assumes, and/or because their disk sample may be mainly drawn from stars for
which the proper motion limit is 0.18'' yr$^{-1}$.  Nevertheless, the results
of this simple simulation provide reason to be skeptical of their claim of the
existence of a retrograde high halo, which clearly {\it can} be influenced by
selection bias of the input sample\footnote{Carney (1999) reported that, after
making a statistical correction for the kinematic bias in the Carney et al.
1996 sample, he also obtained a net prograde rotation even in the high halo.}.

Our large non-kinematically selected sample provides the means to elucidate the
rotational character of the high halo without the effect of an input selection
bias.  Figure 11c presents our results based on sample stars with
[Fe/H] $\le-1.5$. It is found that the stars at small $Z_{max}$ show a net
prograde rotation, in agreement with the results presented in \S 2 and also
with the Carney et al. (1996) result: we obtain $<V_\phi>= 59 \pm 7$
km~s$^{-1}$ for 230 stars at $Z_{max}<2$ kpc.
However, at large $Z_{max}$, the stars exhibit {\it no systematic rotation},
in sharp contrast to the Carney et al. result: we obtain $<V_\phi>=0 \pm 8$
km~s$^{-1}$ for 212 stars at $Z_{max}\ge 4$ kpc.  We note that although
there is a difference in the rotational velocities for the stars close to
and farther from the plane, which may suggest two populations
(accreted and contracted populations), the boundary between them is not obvious.

\subsection{The Global Density Distribution of the Halo}

May \& Binney (1986) discussed the interesting possibility that, on the basis
of Jeans' theorem, the {\it global} structure of the stellar halo can be
recovered from {\it local} kinematic information for a sufficiently large
sample of stars observed in the solar neighborhood.  The theorem states that,
for a well-mixed stellar system, the six-dimensional phase-space distribution
function  of stars, $f({\bf x},{\bf v})$,  can be taken to be a function of the
three isolating integrals of motion $I_i$, $i=1,2,3$, i.e. $f(I_1,I_2,I_3)$.
Within this isolating integral space, the stars constitute a set of fixed
points with no time evolution.  May \& Binney argued that the stars observed in
the solar neighborhood actually occupy a large fraction of this phase space,
and it is hence possible to reconstruct the global structure of the stellar
system from the kinematic data of nearby stars.  Following these strategies,
SLZ developed a maximum likelihood method for recovering a global model of the
halo based on a discrete sum of orbits, and applied it to a sample of 118 local
stars with [Fe/H] $\le-1.5$ selected without kinematic bias.  SLZ found that
the stellar halo at $8<R<20$ kpc may consist of two components -- a main,
nearly spherical component, and an overlapping, highly flattened component.  We
note here that the {\it actual} halo system is unlikely to be in a well mixed
equilibrium state, as we will discuss in the next section.  However, the
relaxation process is very slow compared to the orbital periods of typical
stars, so the Jeans theorem and the above approach based on it are at least
approximately valid.

We now apply the SLZ methodology to the present sample of stars, which is both
larger, and has more accurately determined kinematic information than was
available to SLZ.  To exclude the MWTD stars as much as possible, we select as
representatives of the halo population the stars in our sample with [Fe/H]
$\le-1.8$, a more restrictive abundance cut than the [Fe/H] $ \le-1.5$ used by
SLZ.  We also select a sample of stars with $-1.6<$ [Fe/H] $\le-1$ in order to
examine the characteristics derived for a halo population contaminated by the
MWTD.  To minimize the effects of distance errors we limit our samples to those
stars satisfying $D\le4$ kpc.  We also remove stars with inferred (and possibly
incorrect) extreme space motions ($V_{RF}\le550$ km~s$^{-1}$).  After applying
these cuts, the samples we investigate include $N=359$ stars for [Fe/H]
$\le-1.8$, and $N=302$ stars for $-1.6<$ [Fe/H] $\le-1$ .

The method is summarized as follows (see SLZ for the complete description):
(1) the $N$ sets of isolating integrals $(I_{1,i},I_{2,i},I_{3,i})$,
$i=1,...,N$ are calculated from the observed positions ${\bf x}_i$ and
velocities ${\bf v}_i$ of the stars, within an assumed Galactic potential.
Here, $I_{1}$ is the total orbital energy $E$, $I_{2}$ is proportional
to the square of the angular momentum vector pointing in the $Z$-direction
$I_2=L_z^2/2$ (which measures azimuthal angular momentum), and $I_3$ is the
so-called third integral of motion as defined in the Appendix (see also
equation 15 of SLZ).  For the Galactic potential, we adopt the same St\"ackel
model as in \S 3. (2) At all locations of the stars ${\bf x}_j$, $j=1,...,N$,
the probability density $\rho_i({\bf x}_j)$ of an orbit characterized by
$(E_i,I_{2,i},I_{3,i}$) is calculated for $i=1,...,N$. In other words, we
calculate the $N^2$ matrix $\rho_{ij}$, $i,j=1,...,N$ from knowledge of
integrals and locations of the orbits.  (3) By maximizing the probability that
the star found at ${\bf x}_{j=1}$ is on orbit $i=1$, the star found at ${\bf
x}_{j=2}$ is on orbit $i=2$, and so forth, the orbit weighting factors, $c_i$,
are used to estimate the total density at ${\bf x}$, viz
\begin{equation}
\rho({\bf x}) = \sum_{i=1}^N c_i \rho_i ({\bf x}) \ .
\end{equation}
Thus, equation (3) provides an estimate of the density of the halo stars
at any point ${\bf x}$. The method also permits one to derive the mean
azimuthal velocities as:
\begin{equation}
<V_\phi>({\bf x}) = \frac{1}{\rho({\bf x})}
             \sum_{i=1}^N c_i \rho_i ({\bf x}) V_{\phi,i}({\bf x}) \ .
\end{equation}
We proceed to average the results from equations (3) and (4) over grids of
finite area.  Following SLZ, we define the grids in the meridional plane of the
spheroidal coordinates $(\lambda,\nu)$, which is a suitable choice for
St\"ackel mass models (see the Appendix for more details). The grids are
defined as $\lambda_k=k^2-\alpha$, $k=1,...,30$ and $\nu_l=(\gamma-\alpha)
\cos^2(\theta_l) -\gamma$, $\theta_l=(\pi/2)(l/20)$, $l=0,...20$, as
shown in Figure 4 of SLZ, where $\alpha$ and $\gamma$ are constants.
The spatial resolution of the grids is about 1 kpc.

Figure 12a shows a plot of the reconstructed density distribution at the
Galactic plane (the averaged density over the area at $l=20$), for [Fe/H]
$\le-1.8$ (filled circles) and $-1.6<$ [Fe/H] $\le-1$ (open circles).  As a
comparison, the results using the SLZ sample with [Fe/H] $\le-1.5$ are also
shown (crosses). As in the analysis of SLZ, the density distribution for $R>8$
kpc is well described by a power-law model $\rho \propto R^{\beta}$. For [Fe/H]
$\le-1.8$, we find that the power-law model with exponent $\beta=-3.55\pm0.13$
fits well at all radii beyond $R=8$ kpc, up to the grid point for the largest
radius, $R=35$ kpc.  Note that for the SLZ sample the density at the largest
three radii appears to fall short of the power-law model (probably as a result
of their smaller sample size).  If we omit these outer points, we obtain a fit
to the power-law index $\beta=-3.57\pm0.16$ at $8<R<28$ kpc, which is yet
slightly steeper than the $\beta=-3.29$ result of SLZ.  It is of interest
to note that a power-law model with exponent $\beta\simeq -3.5$ derived
for our sample of field halo stars is similar to the radial density
distribution of the halo globular cluster population derived by Harris (1976)
and Zinn (1985) ($\beta=-3.5$) and by Carney, Latham, \& Laird (1990)
($\beta=-3.0$).
A similar density distribution, but with a slightly shallower slope, has been
found for field RR Lyrae stars by Saha (1985) ($\beta\simeq-3$) and Hawkins
(1984) ($\beta \simeq-3.1$).  Preston et al. (1991) combined counts of
RR Lyrae stars and FHB stars in several fields to obtain the exponent
$\beta = -3.5 \pm 0.3$.
We see, in our reconstructed density distribution, no clear evidence
for a break in the density distribution at $R=20-25$ kpc as was detected
in the number counts of globular clusters and RR Lyraes.
For the subsample of stars with $-1.6<$ [Fe/H]
$\le-1$, we obtain $\beta=-3.47\pm0.18$ for $8 < R < 25$ kpc, thus
contamination from the MWTD has little effect.  Below $R=8$ kpc, the density
distributions of all three samples clearly deviate from a single power-law
model, a result which is likely caused by incomplete representation, in the
solar neighborhood, of stars for which apocentric radii are below $R=R_\odot$.

Figure 12b shows the mean azimuthal velocities, $<V_\phi>$, for the same three
samples of stars, projected onto the Galactic plane. The value of $<V_\phi>$ is
nearly zero for $R>10$ kpc, but there is a signature of {\it increasing}
$<V_\phi>$ with decreasing $R$. In particular, for the sample of stars with
$-1.6<$ [Fe/H] $\le-1$, $<V_\phi>$ rises rather discontinuously at $R\simeq 10$
kpc, which may correspond to the radial limit of the rapidly rotating
thick-disk component.

Figure 13 is a plot of the inferred global density distribution in the $(R,Z)$
plane, in the form of equidensity contours. The lack of stars at small $R$ and
large $Z$ (which gives rise to the ill-formed contour levels in this portion of
the diagram) is a consequence of the small probability that stars in the Galaxy
that explore such a region are represented in the solar neighborhood (as
argued in SLZ).  Other than in this region, the inferred density distribution
based on the 359 stars with [Fe/H] $\le-1.8$ (panel b) appears to be very
similar to that which SLZ obtained from their 118 stars (panel a):  The outer
part of the halo, at $R>15$ kpc, is round, in good agreement with inferences
based on star counts (see Freeman 1987 and references therein; Preston et al.
1991).

While we reproduce the general sense of the SLZ results with our much larger
sample, there are notable differences in the details.  SLZ argued that there is
a clear indication that the halo, at any given radius, consists of {\it both} a
main, nearly spherical component and an {\it overlapping}, highly flattened
component.  There no clear evidence for this result in the density
reconstruction based on our new data.  To further examine this point, we have
fit elliptical contours to the reconstructed density maps after, following SLZ,
{\it omitting} the data points near the Galactic plane.  Specifically, we
obtain fits to ellipses of major axis $a$, and axial ratio $q$, over the polar
angle $40^\circ<\theta<80^\circ$. Residuals to the fits obtained to these
ellipses as a function of polar angle are shown in Figure 14a.  Thick solid and
dotted lines correspond to the fits with $a=10.5$ kpc and $q=0.70$ and with
$a=13.5$ kpc and $q=0.51$, respectively, for our sample with
[Fe/H] $\le-1.8$.  For comparison, we show the corresponding results using the
SLZ sample fit over $30^\circ<\theta<80^\circ$ (thin solid and dotted lines).
It follows that, while we reproduce the SLZ result that there is a large
density excess near the Galactic plane when using {\it their} sample, it is not
evident in {\it our} sample -- an additional flat component is not required for
the fitting.

Based on the above result, we proceed to make a fit including the density data
near the plane, with a single value for the axial ratio $q$ at each major axis
$a$, i.e., without taking into account an additional flat component having
small $q$.  The change of our estimate of $q$ as a function of radius
is shown in Figure 14b.  For the sample of stars with [Fe/H] $\le-1.8$, the
density distribution in the outer part of the halo, $R \sim 20$ kpc, is quite
round.  However, the axial ratio $q$ appears to decrease with decreasing $R$
over $15<R<20$ kpc, and the inner part, at $R<15$ kpc, exhibits $q \sim 0.65$.
Thus, the halo can be described as nearly spherical in the outer part and
highly flattened in the inner part, instead of the overlap of both components
at all locations in the halo.  This result is in good agreement with previous
studies of the distribution of RR Lyrae (Hartwick 1987; Layden 1995) and FHB
stars (Preston et al. 1991; Kinman, Suntzeff, \& Kraft 1994), as well as with
the flattening of the inner halo reported by Larsen \& Humphreys (1994) based
on counts of F- and G-type stars.  It remains an open question as to whether or
not there exists a distinct boundary between the outer spherical halo and the
inner flattened halo.

As seen in Figure 13c, the density distribution in the inner part of the halo
for stars of intermediate abundance ($-1.6<$ [Fe/H] $\le-1$) appears to be more
flattened than is the case for stars with [Fe/H] $\le-1.8$.  This result is
reflected in the ellipse fits for these stars shown in Figure 14b, which
indicate that $0.5<q<0.6$ at $R<12$ kpc. This may be due to the contribution of
the (by definition) flattened, MWTD population with a finite radial scale
length. We note that our results imply the decrease of axial ratios at $R>17$
kpc. This may be an artifact due to the small numbers of stars employed at such
large radii, so further analysis using much larger samples is necessary.

\section{Kinematic Substructure of the Halo in the Solar Neighborhood}

If the halo of the Galaxy was assembled from the merging and/or accretion of
small subgalactic clumps, as argued by SZ, then one might hope to find
signatures of those events, even now, in the form of kinematic substructures,
because the mixing of phase space for such stars is expected to be incomplete
(Helmi \& White 1999).  The Sagittarius dwarf galaxy is an ongoing merging
event at the current epoch (Ibata, Gilmore, \& Irwin 1994), and the Magellanic
Clouds may ultimately follow the similar fate, as they lose energy via
dynamical friction (Tremaine 1976).  Signatures of past merging events in the
halo were reported by Majewski et al. (1994) in their {\it in situ} sample of
the stars at about 4.5 kpc above the Galactic plane.  Helmi \& White (1999)
argue that the reported clumpiness in the velocity distribution of the halo
stars from Majewski et al. is actually a superposition of two individual
streams of stars, possibly arising from a common progenitor.

Clear ``fossil evidence'' in the solar neighborhood for a previous merger of
the Galaxy with what may have been a ``Seale \& Zinn fragment'' has recently
been discovered by Helmi et al.  (1999, hereafter HWdZZ).  These authors
examined a subset of the stars in samples from our previous work
(BSL; CY), and identified a statistically significant clumping of stars in the
angular momentum diagram $L_z$ vs $L_\perp =(L_x^2+L_y^2)^{1/2}$.  The
substructure identified by HWdZZ consists of 7 stars (in a sample of 97 stars
with [Fe/H] $\le-1.6$ and $D < 1$ kpc), or 12 stars (in a sample of 275 stars
with [Fe/H] $\le-1$ and $D < 2.5$ kpc).  HWdZZ suggest, based on the observed
numbers of stars in this clump, that roughly 10\% of the halo stars outside
the solar radius may have arisen from a single coherent object with a
total mass of about $10^8$ M$_\odot$, disrupted during the process of
halo formation.

We now consider the HWdZZ result based on our revised catalog.  Figure 15 shows
the angular momentum diagram of HWdZZ as populated by the stars of our present
sample within 2.5 kpc of the Sun. Panels (a) and (b) are for stars in the
abundance ranges [Fe/H] $\le-1.6$ and $-1.6<$ [Fe/H] $\le-1$, respectively.  In
both of these abundance ranges there exists a clump of stars in the region that
HWdZZ pointed out, at $L_\perp \sim 2200$ kpc~km~s$^{-1}$ and $L_z \sim 1200$
kpc~km~s$^{-1}$.  In addition to this clump, we identify a possible ``trail''
(in angular momentum space) which appears to connect the clump and the high
$L_z$ region, most clearly evident among the higher abundance stars shown in
panel (b).  For the purposes of this discussion, we define the ``clump'' region
to be comprised of stars with $2100 < L_\perp < 2600$ kpc~km~s$^{-1}$ and $800
< L_z < 1500$ kpc~km~s$^{-1}$ (solid box), and the ``trail'' region to be
comprised of stars having $1250 < L_\perp < 2000$ kpc~km~s$^{-1}$ and $1200 <
L_z < 2000$ kpc~km~s$^{-1}$ (dotted box), respectively. We include BPS CS
22876-0040 as a ``trail'' member (triangle in Fig.15a), which is somewhat
outside the region defined above, $(L_z,L_\perp)=(1037,1672)$ kpc~km~s$^{-1}$,
because this star exhibits similar orbital motions to the ``trail'' stars
examined below.

We note that the angular momentum, $L_\perp$, is not an exact integral of
motion in the currently adopted Galactic potential (which consists of a disk
and halo component), thus its use may not be generally appropriate for the
study of kinematic substructures in the halo.  Rather, the so-called third
integral, $I_3$, which is related to $L_\perp$ in a non-spherical potential,
should be used.  Unfortunately, no general analytic expression exists for
$I_3$.  However, the St\"ackel form of the currently adopted potential allows
one to estimate $I_3$ in an explicit manner (de Zeeuw 1985; de Zeeuw, Peletier,
\& Franx 1986; Dejonghe \& de Zeeuw 1988), as well as other integrals, such as
the orbital energy $E$ and the angular momentum $L_z$.  Expressions for
these three integrals are described in the Appendix.

For a spherically symmetric potential, the quantity $(2I_3)^{1/2}$ is
equivalent to $L_\perp$.  Figure 16 is a plot of $(2I_3)^{1/2}$ vs. $L_z$
(panel a), and $|E|$ vs $L_z$ diagram (panel b), respectively, for the 723
stars from our sample with [Fe/H] $\le-1$ and $D\le 2.5$ kpc.  Filled and open
circles denote the stars in the ``clump'' and ``trail'' regions of Figure 15,
respectively.  We note that one star in the ``clump'' region of Figure 15
(HD~214161) exhibits quite different orbital parameters than the others (see
below), so this star is drawn as a cross in Figure 16.  The fact that we see
such similar structures in Figures 15 and 16 implies that, along the orbits of
the stars constituting the ``clump'' and ``trail'', the gravitational potential
can be regarded as nearly spherical -- the member stars spend the majority of
their orbits far from the Galactic plane, where the effect of the disk
potential is modest.  Also, panel (b) suggests that the orbital energies $|E|$
of the stars in the ``clump'' are confined to the narrow range near $|E| \simeq
10^5$ km$^2$ s$^{-2}$, whereas the stars in the ``trail'' have a rather diverse
range of $|E|$.

Another choice of integrals in the current mass model are the so-called action
integrals, ${\bf J}$. Action integrals are adiabatic invariants, and thus
remain unchanged even if the orbital energy $E$ changes via gravitational
interaction, provided the time scale for the interaction is sufficiently long
compared to the orbital period.  One of the actions is $J_\phi$, which is
equivalent to $L_z$ in an axisymmetric potential.  Other suitable actions in
the current St\"ackel potential, defined in terms of the spheroidal coordinates
$(\lambda,\nu)$, are $(J_\lambda,J_\nu)$ (see the Appendix for complete
definitions).  Panels (c) and (d) of Figure 16 show the distribution of our
sample stars in the $J_\nu$ vs $J_\lambda$ and $J_\nu$ vs $L_z$ diagrams,
respectively. It can been seen that all of the ``clump'' stars, except for the
star HD 214161 $(J_\lambda=5058,J_\nu=1223)$, exhibit a clear clumpiness in
action space, whereas the ``trail'' stars show a broad distribution in $J_\nu$,
but with a rather narrow distribution in $J_\lambda$ [with the exception of the
star CS~Ser $(J_\lambda=6957,J_\nu=587)$].

From these integrals of motion, we conclude that the ``clump'' consists of only
9 stars, instead of 12 stars in the HWdZZ result, and the ``trail'' consists
of 9 stars. The kinematic quantities for these stars are listed in Table 6. The
orbit of the ``clump'' is characterized by $Z_{max}\sim 16$ kpc, $r_{ap}\sim
20$ kpc, and $r_{pr}\sim 7$ kpc, which are in good agreement with the HWdZZ
result.  Note that even though we have tripled the numbers of stars considered
in our present sample (728 stars), relative to that of the sample examined by
HWdZZ (275 stars), the number of detected clump members has {\it not}
increased. Thus, HWdZZ's conclusion that as much as one-tenth of the halo stars
presently located in the solar neighborhood originates from a single object may
not apply.  We have not considered, as of yet, the impact of the ``trail''
stars on this argument.

Figure 17 shows the metallicity distribution of the stars in the ``clump''
(solid histogram) and in the ``trail'' (dotted histogram) features.  The two
features seem to share a similar metallicity distribution, though the number of
stars involved is still too small for any definite conclusion to be reached.
It might be tempting to suggest that the ``trail'' was formed via the tidal
interaction of the precursor object of the ``clump'' with the Galactic
potential, with the ``trail'' stars gaining some of the orbital angular
momentum lost by the precursor object.  Such an interaction may have proceeded
in a rapid manner, being comparable to the orbital period of the object, so
that the actions of the ``trail'' differ from those of the ``clump''.  Progress
on evaluation of this picture will require detailed numerical simulations,
which we are presently investigating.

\section{Discussion and Conclusion}

We have analyzed both the local and global kinematics of 1203 metal-poor stars
in the Galaxy with [Fe/H] $\le-0.6$, based on a large, revised, catalog of
stars selected without kinematic bias (Paper II).  All of these stars have
available distance estimates, radial velocities, proper motions, and abundance
estimates over the full applicable range in the Milky Way.  This is
the largest non-kinematically selected sample yet assembled, so the derived
kinematic properties are the least affected by systematics as well as
statistical fluctuations. We summarize our results below, and discuss them in
the context of the formation of the Galaxy.

\subsection{Summary of the Results}

The local kinematics of the halo population, based on the stars with
[Fe/H] $ \le-2.2$ and $|Z|<1$ kpc, are characterized by a radially elongated
velocity ellipsoid $(\sigma_U,\sigma_V,\sigma_W)=(141 \pm 11,106 \pm 9,94 \pm
8)$ km~s$^{-1}$ and a small prograde rotation $<V_\phi>=30 \sim 50$ km~s$^{-1}$
(assuming $V_{LSR}=220$ km~s$^{-1}$).  When additional halo stars at larger
$|Z|$ are taken into account, the velocity ellipsoid remains essentially
unchanged, but $<V_\phi>$ exhibits a marked decrease (Figs. 2 and 3). We find
no evidence for an increase of $\sigma_W$ at the lowest abundances, as had been
previously suggested.  At higher metallicities, the stars in our sample exhibit
disk-like kinematics, and a higher mean rotation.  Specifically,
for stars in the abundance interval $-0.7\le$ [Fe/H] $<-0.6$ and with $|Z|<1$
kpc, we have obtained $(\sigma_U,\sigma_V,\sigma_W) = (46 \pm 4,50 \pm 4,35 \pm
3)$ km~s$^{-1}$ and $<V_\phi>=200$ km~s$^{-1}$, which characterize
the kinematic parameters of the thick disk. We have also confirmed previous
results that there exists a remarkable discontinuity of the rotational
properties of the Galaxy at [Fe/H] $\simeq -1.7$ (Fig. 3).

Analysis of a large sample of non-kinematically selected stars provides clear
evidence, supporting earlier suspicions based on much smaller samples, that
there exists {\it no correlation between metal abundances and orbital
eccentricities} for metal-poor stars of the Milky Way (Fig. 5). Even at the
lowest abundances explored in our sample, [Fe/H] $\le-2.2$, about 20\% of the
stars have $e<0.4$. In addition, there is a small concentration of high-$e$
stars at [Fe/H] $\sim-1.7$, which is possibly responsible for the near zero
$<V_\phi>$ at the same metallicity. We found that
the fraction of the low-eccentricity stars with
[Fe/H] $\le-2.2$ remains the same, even as one changes the range of $|Z|$ (Fig.
6a), so such stars belong to the halo, not the MWTD component.
On the other hand, stars in intermediate abundance ranges above $-2.2$ dex
exhibit a decrease of low-$e$ stars with increasing $|Z|$, and the fraction of
such stars appears to converge at high $|Z|$ to that found for [Fe/H] $\le-2.2$
(Fig. 6b). Both a KS test and a Monte Carlo simulation enable a determination
of the structural parameters of the disk component in these abundance ranges.
Specifically, the fraction of the disk component is about 30\% for
$-1.7<$ [Fe/H] $\le-1$, but is less than 10\% for more metal-poor ranges (Fig.
8).

The global kinematics of the halo stars are summarized as follows. In contrast
to the claims of Majewski (1992), and Carney et al. (1996), stars in our sample
do {\it not} show a net retrograde rotation at large $Z_{max}$, but rather
exhibit a near zero systematic rotation (Fig. 10).  The difference between our
result and that of Carney et al. (1996) probably arises from the (unavoidable)
kinematic bias inherent in their sample selection criteria.  The observed
decrease of $<V_\phi>$ with increasing $Z_{max}$ is continuous, so that it is
not possible to conclude that the inner ``contracted'' population (with a
positive $<V_\phi>$) is distinct from an outer ``accreted'' population, based
on the rotational properties of the metal-poor stars alone. Our analysis of the
global density distribution of halo stars, based on the reconstruction
method developed by SLZ, confirms SLZ's conclusion that the outer halo is quite
round (Fig.  12). However, we see no evidence of an overlapping flattened
component in addition to the main, nearly spherical one, as was claimed by SLZ.
Rather, the density distribution of the halo is better described as nearly
spherical in the outer region (beyond $R=15-20$ kpc) and highly flattened in
the inner region.

We have confirmed a recent detection of kinematic substructure in the solar
neighborhood by HWdZZ, based on a small number of stars which cluster together
in the halo angular momentum diagram.  We have also found an additional
elongated ``trail'' which appears to connect between HWdZZ's ``clump'' and the
high $L_z$ region (Fig. 14).  Further analysis, using several integrals of
motion for the ``clump,'' does not result in a dramatic increase in the numbers
of stars associated with it, even though the total number of our sample stars
is three times as large as that available to HWdZZ.

\subsection{Implications for the Formation of the Galaxy}

The local and global kinematics of metal-poor stars provide valuable clues for
understanding the formation process of the halo and thick disk components in
the Galaxy, as well as in disk-type galaxies in general.

If the primordial collapse from the halo to the disk occurred in a monolithic
manner, starting from an overdense homogeneous spheroid, one might expect (as
predicted by the ELS model) to observe a continuous increase of $<V_\phi>$ for
the stars born from the infalling gas, as well as a continuous decrease of
their orbital eccentricities with increasing [Fe/H] as the spheroid spins up
in order to conserve angular momentum.  The fact that we observe no correlation
between [Fe/H] and $e$, and basically no change of $<V_\phi>$ with abundance
for stars with [Fe/H] $\le-1.7$ conflicts with this scenario.  The lack of an
abundance gradient in the halo stars (Carney et al. 1990; CY) is also difficult
to interpret in this context.  The outer halo, if formed from a monolithic
collapse, might be expected to be dominated by radially elongated motions of
the stars, but this is actually opposite to the inferred tangentially
anisotropic velocity ellipsoid at large distance from the Sun (see
Sommer-Larsen et al. 1997). We also note that a small portion of the metal-poor
stars having [Fe/H] $\sim-1.7$ may have been formed from the infalling part of
gas, so as to explain both the nearly zero $<V_\phi>$ and the excess number of
high-$e$ stars found at [Fe/H] $\sim-1.7$.

If the halo is assembled from merging and/or accretion of numerous fragments
falling into the Galaxy (SZ), one might expect little or no correlation
between kinematic and chemical properties, as each fragment has its own
chemical history, and the merging process may proceed in a chaotic manner.  Our
results for the $<V_\phi>$ vs. [Fe/H] and [Fe/H] vs. $e$ relations are
basically in agreement with this scenario.  The SZ scenario is also consistent
with a several Gyr age spread in globular clusters in the outer halo (see,
e.g., Rosenberg et al. 1999), and even in field stars (Schuster \& Nissen
1989), because the initiation and duration of star formation may not be
coherent from fragment to fragment\footnote{Harris et al. (1997) showed that
the most metal-poor globular clusters, such as M92, have essentially the same
age everywhere in the halo. As they argued, this result could also be explained
within the precepts of the SZ scenario if all of the ``SZ fragments'' began
building the first generation of clusters in the same time period.}.  However,
the original SZ scenario seems unlikely to explain our observed vertical
gradient of $<V_\phi>$ for halo stars, as well as the highly flattened density
distribution of the inner halo, in contrast to the nearly spherical outer halo.
Totally incoherent, chaotic merging of SZ fragments would not be expected to
produce these ``internal'' kinematic structures in the halo. It is also unclear
as to how the rapidly rotating disk component subsequently formed out of the
aftermath of merging (see also Freeman 1996).

SZ first suggested that at least the inner part of the halo may have undergone
a coherent contraction in a manner similar to the ELS hypothesis, an idea
which has been invoked by subsequent workers to explain the duality of the
density, kinematics, and ages of the halo field stars (e.g., SLZ; Norris 1994;
Carney et al. 1996; Sommer-Larsen et al. 1997), as well as the age difference
between outer and inner globular clusters (Zinn 1996; Rosenberg et al. 1999).
This sort of hybrid picture, combining aspects of both the ELS and SZ
scenarios, proposes that the outer halo is made up from merging and/or
accretion of subgalactic objects, such as dwarf-type satellite galaxies,
whereas the inner part of the halo has undergone a dissipative contraction on
relatively short timescales.  This hybrid model might explain our
identification of the inner, flattened, slowly rotating component of the halo
with a finite spatial gradient in $<V_\phi>$.

An alternative hypothesis to explain an observed ``duality'' of the Galactic
halo relies on the existence of a thick-disk population of stars even at rather
low abundances (MFF; Norris 1994; BSL). If stars with disk-like kinematics have
a metallicity distribution which extends below [Fe/H] $=-2$, then a finite
fraction of their orbits would be characterized by low $e$, as found in our
present investigation.  One possible origin of this MWTD component may be the
heating of the pre-existing thin disk triggered by the dissipationless merging
of a satellite falling into the disk (Quinn, Hernquist, \& Fullagar 1993).
Under this hypothesis, the currently observed thin disk, with a vertical scale
height of $\sim 350$ pc, could only have formed {\it after} the merging event
was completed.  However, our finding that few thick-disk stars exist with
[Fe/H] $\le-1.7$, and no observed increase of $\sigma_W$ with decreasing
[Fe/H], belies the existence of a dynamically hot, proto-disk population at the
lowest abundances (see also Norris 1994; Ryan \& Lambert 1995; Twarog \&
Anthony-Twarog 1996).  Furthermore, following the results presented in \S 4.2,
we see no evidence for an overlapping flattened component of the halo in
addition to a nearly spherical component.  An indication that there might exist
a significant vertical gradient in $<V_\phi>$ for [Fe/H] $\le-1.7$, compared
with a much smaller gradient observed for the thick disk itself, also conflicts
with this hypothesis. Thus, we conclude that the formation of the inner
flattened halo possibly involves a dissipative contraction, not a
dissipationless heating of the proto-disk.

If a hybrid halo formation picture, based on dissipationless merging in the
outer halo and dissipative contraction in the inner halo, applies, the question
arises as to whether there is a clear boundary distinguishing the two regions.
The results presented in \S 4 suggest no clear distinction between the outer
and inner regions of the halo, at least as seen from inspection of the
$<V_\phi>$ vs. $Z_{max}$ relation, and the inferred globafl density distribution
of the halo. Furthermore, there presently exists no reasonable theoretical
explanation for the existence of two distinct populations of stars in the halo.
Thus, our current analysis implies that both dissipationless and dissipative
processes in the outer and inner halo, respectively, may have occurred more or
less in a simultaneous manner.

We now ask whether the above hybrid scenario is a natural consequence of the
currently favored theory of galaxy formation based on hierarchical assembly of
cold dark matter (CDM) halos (e.g., Peacock 1999).  The CDM model postulates
that initial density fluctuations in the early Universe have larger amplitudes
on smaller scales.  Thus, the initially overdense regions that end up forming
large galaxies such as our own contain large density fluctuations on
subgalactic scales.  As a protogalaxy collapses from the general cosmological
expansion, these small-scale fluctuations develop into numerous clumps of CDM
particles, into which the interstellar gas falls from gravitational attraction.
The protogalaxy is thus made up of numerous clumps comprised of a mixture of
primordial gas and dark matter, interacting with one another via their mutual
gravitational attraction.  According to numerical simulations by Steinmetz \&
M\"uller (1994; 1995) and Bekki \& Chiba (1999), most of the metal-poor stars
which presently occupy the outer halo of our Galaxy form in these local,
small-scale density fluctuations.  Once star formation initiates, the gas
inside of these small fragments quickly escapes due to energy feedback from
supernovae.  Later, these clumps begin to merge with one another, and the
aftermath of these essentially dissipationless merging processes exhibits a
nearly spherical density distribution with no abundance gradient.

The subsequent evolution of the system may be described in the following way
(Bekki \& Chiba 1999). As a consequence of the merging of low-mass fragments, a
smaller number of more massive clumps develops -- within each of these merged
clumps one expects to find previously formed metal-poor stars as well as newly
born stars.  These large clumps continue to accrete gas from their immediate
surroundings.  These clumps gradually move toward the central region of the
system due to both dynamical friction, and dissipative merging with smaller
clumps.  Then, when the last merging event between the largest clumps occurs,
the stars which have been confined inside the clumps are disrupted and spread
over the inner part of the halo, whereas a large fraction of the disrupted gas
appears to end up in the center of the Galaxy and may form a bulge.  As a
consequence, the inner part of the halo should have a flattened density
distribution with a finite prograde rotation, as reported here, and its angular
momentum distribution may be similar to that of the bulge (Wyse \& Gilmore
1992). Also, the stars born from this infalling stage of gas may explain the
existence of high-$e$ stars at [Fe/H] $\sim-1.7$.  The simulations conducted to
date imply that the thick disk component is partially composed of debris stars,
but it is mainly made from diluted gas which has been accreted from the outer
part of the halo (Sommer-Larsen et al.  1997).  Therefore, although more
detailed simulation work is required, CDM models appear to reproduce, at least
qualitatively, the overall kinematic properties of the metal-poor stars via
both dissipationless merging in the outer halo, and dissipative merging in the
inner halo.

It is unknown whether or not evidence for merging events of CDM clumps during
the early evolution of the Galaxy might be still preserved as kinematic
substructures at the current epoch.  Within the currently available precision
of space velocities for stars in our sample, typically of the order of 10 to 20
km~s$^{-1}$, the main body of the halo appears to be well mixed in phase space;
higher precision measurements of proper motions by the planned astrometric
satellites missions (e.g., {\it FAME, SIM, GAIA}) may be able to disentangle
this complex mixture of halo stars (Helmi, Zhao \& de Zeeuw 1999).
Alternatively, the confirmed kinematic clumping of halo stars presented in \S 5
may originate from the recent accretion of a satellite galaxy, which has fallen
into the Galaxy after the major part of the halo was formed.

Firmer conclusions on the formation of the Galaxy require the assembly and
analysis of still larger numbers of stars with accurate distances and proper
motions, especially at larger distances from the Sun.  Exploration along this
line is now in progress.  More elaborate numerical modeling of the formation of
large spiral galaxies such as the Milky Way is also needed in order to clarify
the physical processes that lead to the currently observed dynamics and
structure of the halo and disk components.  It is of particular importance to
model and understand the chemo-dynamical evolution of the system of subgalactic
fragments in the course of the Galaxy's collapse.  Once a fundamental
understanding of the formation and evolution of {\it our} Galaxy is
established, it will then be possible to obtain additional insights into
formation of disk-type galaxies in general, by combining our refined picture
with the rapidly growing observational database of young galaxies becoming
available in the deep realm of the Universe.

\acknowledgments

We are grateful to the referee, Bruce Carney, for his careful review of the
paper and a number of thoughtful suggestions.
MC acknowledges partial support from Grants-in-Aid for Scientific
Research (09640328) from the Ministry of
Education, Science, Sports and Culture of Japan.  TCB acknowledges partial
support for this work from grant AST 95-29454 from the National Science
Foundation.

\clearpage
\appendix
\section{The St\"ackel Potential and Integrals of Motion}

We briefly describe the properties of the St\"ackel potential adopted in this
work, and present expressions for the associated integrals of motion. For more
details, see, e.g., de Zeeuw (1985), Dejonghe \& de Zeeuw (1988), and SLZ.

We define spheroidal coordinates $(\lambda,\phi,\nu)$, where $\phi$ corresponds
to the azimuthal angle in cylindrical coordinates $(R,\phi,Z)$, and $\lambda$
and $\nu$ are the roots for $\tau$ of
\begin{equation}
\frac{R^2}{\tau+\alpha} + \frac{Z^2}{\tau+\gamma} = 1 \ ,
\end{equation}
where $\alpha$ and $\gamma$ are constants, giving
$-\gamma\le\nu\le-\alpha\le\lambda$. The coordinate surfaces are spheroids
$(\lambda=const.)$ and hyperboloids of revolution $(\nu=const.)$ with the
$Z$-axis as the rotation axis, where the focal distance
$\Lambda=(\gamma-\alpha)^{1/2}$ fixes the coordinate system.

The gravitational potential of the St\"ackel type is then written as
\begin{equation}
\varphi(\lambda,\nu) = -
 \frac{(\lambda+\gamma)G(\lambda)-(\nu+\gamma)G(\nu)}{\lambda-\nu} \ ,
\end{equation}
where $G(\tau)$ is an arbitrary function. In this work, $G(\tau)$ is the
sum of $G_{disk}(\tau)$ from the disk and $G_{halo}(\tau)$ from the massive
dark halo. Following SLZ, we adopt the oblate perfect spheroid for
$G_{disk}(\tau)$ and the $s=2$ model of de Zeeuw, Peletier, \& Franx (1986)
for $G_{halo}(\tau)$.

The Hamiltonian, $H$, per unit mass, for motion in this potential
$\varphi(\lambda,\nu)$ is
\begin{equation}
H = \frac{p_\lambda^2}{2P^2} + \frac{p_\phi^2}{2R^2}
  + \frac{p_\nu^2}{2Q^2} + \varphi(\lambda,\nu) \ ,
\end{equation}
where $P$ and $Q$ are the metric coefficients of the spheroidal coordinates,
given by
\begin{equation}
P^2=\frac{\lambda-\nu}{4(\lambda+\alpha)(\lambda+\gamma)}, \
Q^2=- \frac{\lambda-\nu}{4(\nu+\alpha)(\nu+\gamma)}  \ ,
\end{equation}
and $p_\lambda$, $p_\phi$, and $p_\nu$ are the conjugate momenta to
$\lambda$, $\phi$, and $\nu$, respectively,
\begin{equation}
p_\lambda=P^2\dot{\lambda}=Pv_\lambda,\ p_\phi=R^2\dot{\phi}=Rv_\phi,\
p_\nu=Q^2\dot{\nu}=Qv_\nu \ .
\end{equation}
The velocities $v_\lambda$, $v_\phi$, and $v_\nu$ at a point
$(\lambda,\phi,\nu)$ are the components of the velocity ${\it\bf v}$ along
the orthogonal axis defined locally by the spheroidal coordinate system.

The three integrals of motion, $|E|$, $I_2$, and $I_3$, are defined as
\begin{eqnarray}
|E| &=& -H \\
I_2 &=& \frac{L_z^2}{2} \\
I_3 &=& \frac{1}{2}(L_x^2+L_y^2) + \Delta^2
   \left[ \frac{1}{2}v_z^2 - Z^2 \frac{G(\lambda)-G(\nu)}{\lambda-\nu}
   \right] \ ,
\end{eqnarray}
and the action integrals $J_\lambda$, $J_\phi$, and $J_\nu$ are defined as
\begin{eqnarray}
J_\lambda &=& \frac{1}{2\pi} \oint p_\lambda d\lambda =
         \frac{2}{\pi} \int^{\lambda_2}_{\lambda_1} p_\lambda d\lambda \\
J_\phi    &=& \frac{1}{2\pi} \oint p_\phi d\phi = L_z \\
J_\nu     &=& \frac{1}{2\pi} \oint p_\nu d\nu =
         \frac{2}{\pi} \int^{\nu_0}_{-\gamma} p_\nu d\nu \ ,
\end{eqnarray}
where ($\lambda_1,\lambda_2)$ and $\nu_0$ are the turning points of the orbit,
defined as the values for which $v_\lambda=0$ and $v_\nu=0$, respectively,
and $\nu=-\gamma$ defines the equatorial plane. For the evaluation of
$J_\lambda$, we have taken four times the integrals from $\lambda_1$ to
$\lambda_2$, to maintain symmetry between $J_\lambda$ and $J_\nu$ and
ensure continuity of the actions across transitions from one orbital
family to another (de Zeeuw 1985).

\clearpage

\centerline{\psfig{file=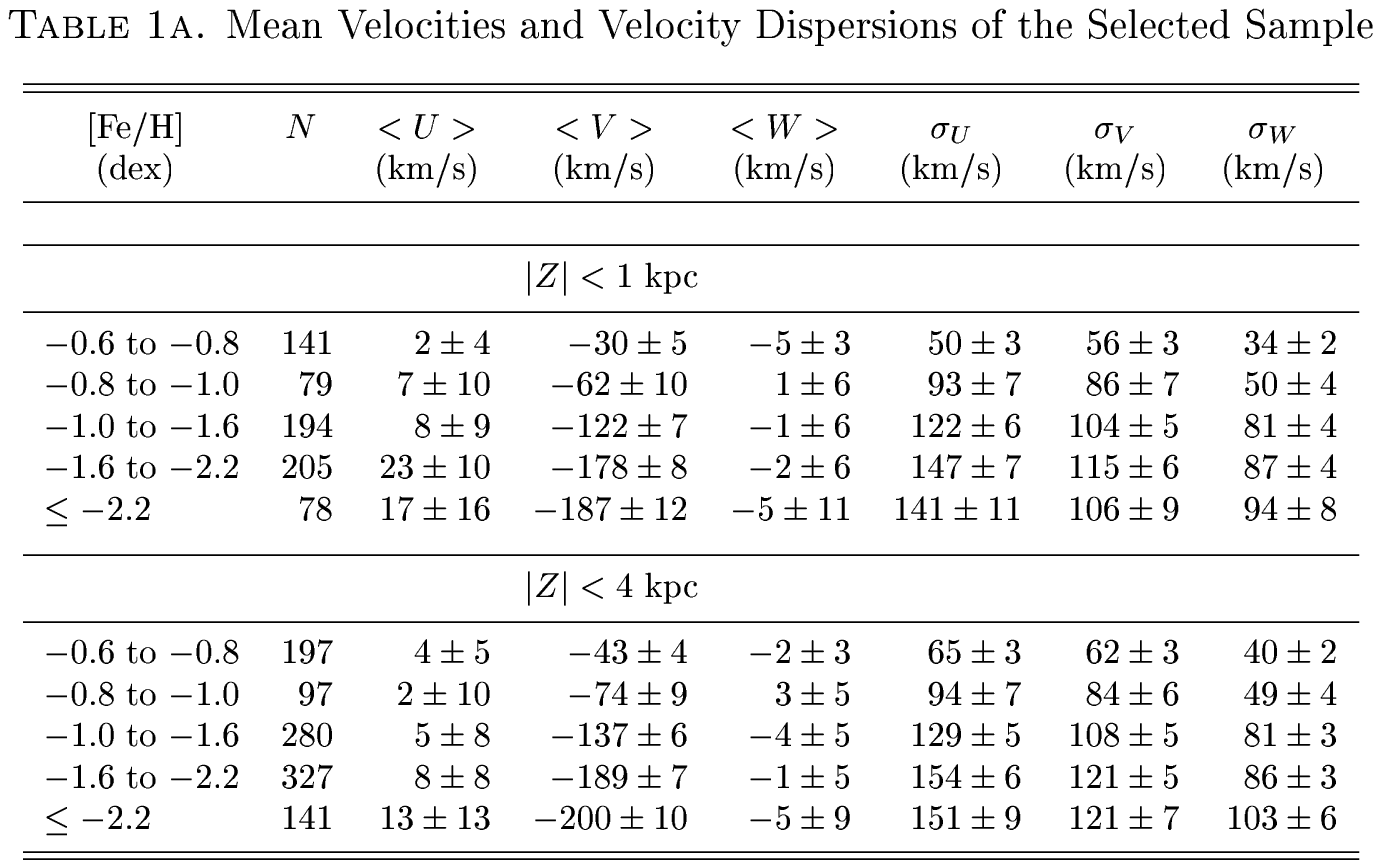}}
\centerline{\psfig{file=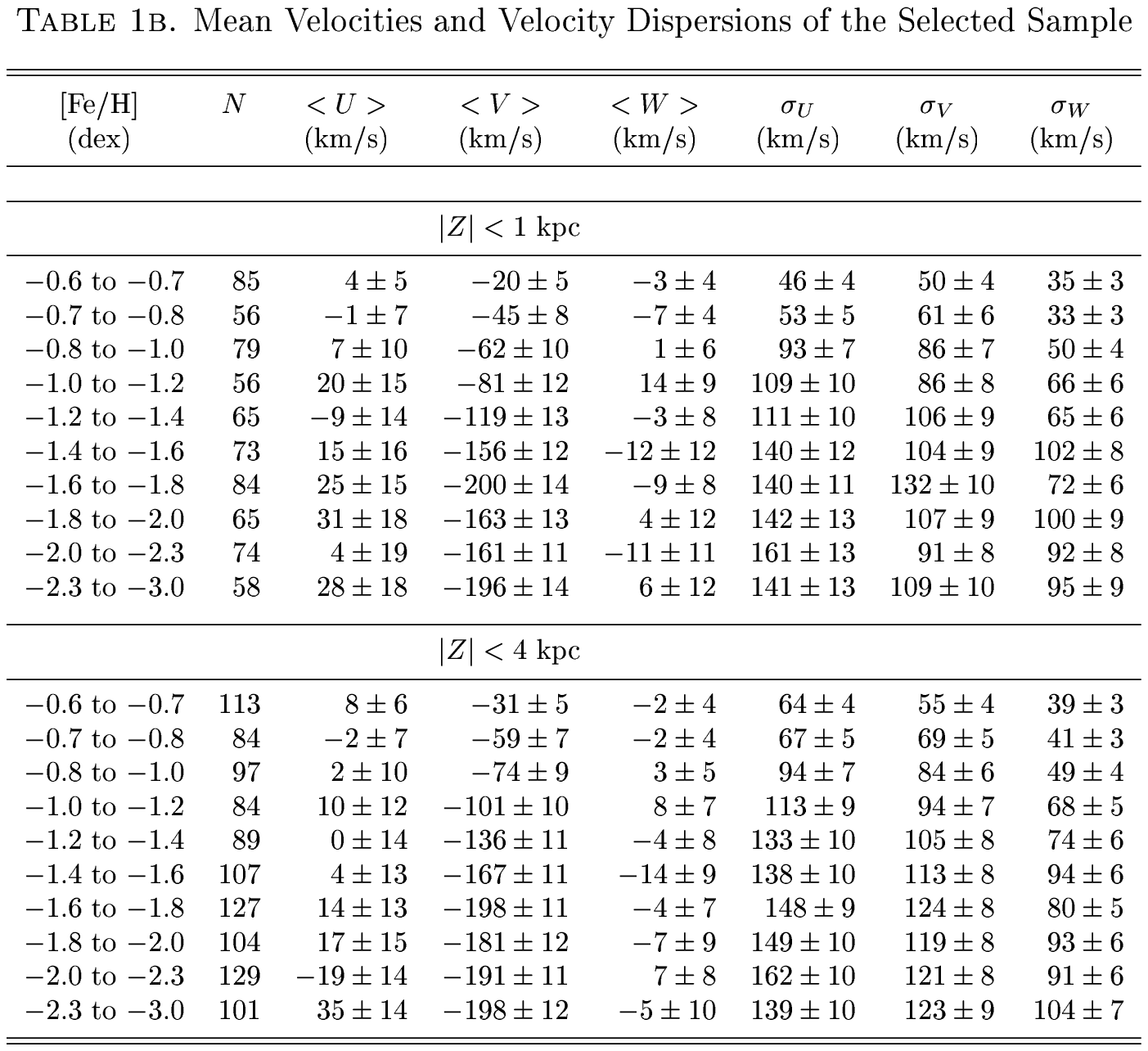}}
\centerline{\psfig{file=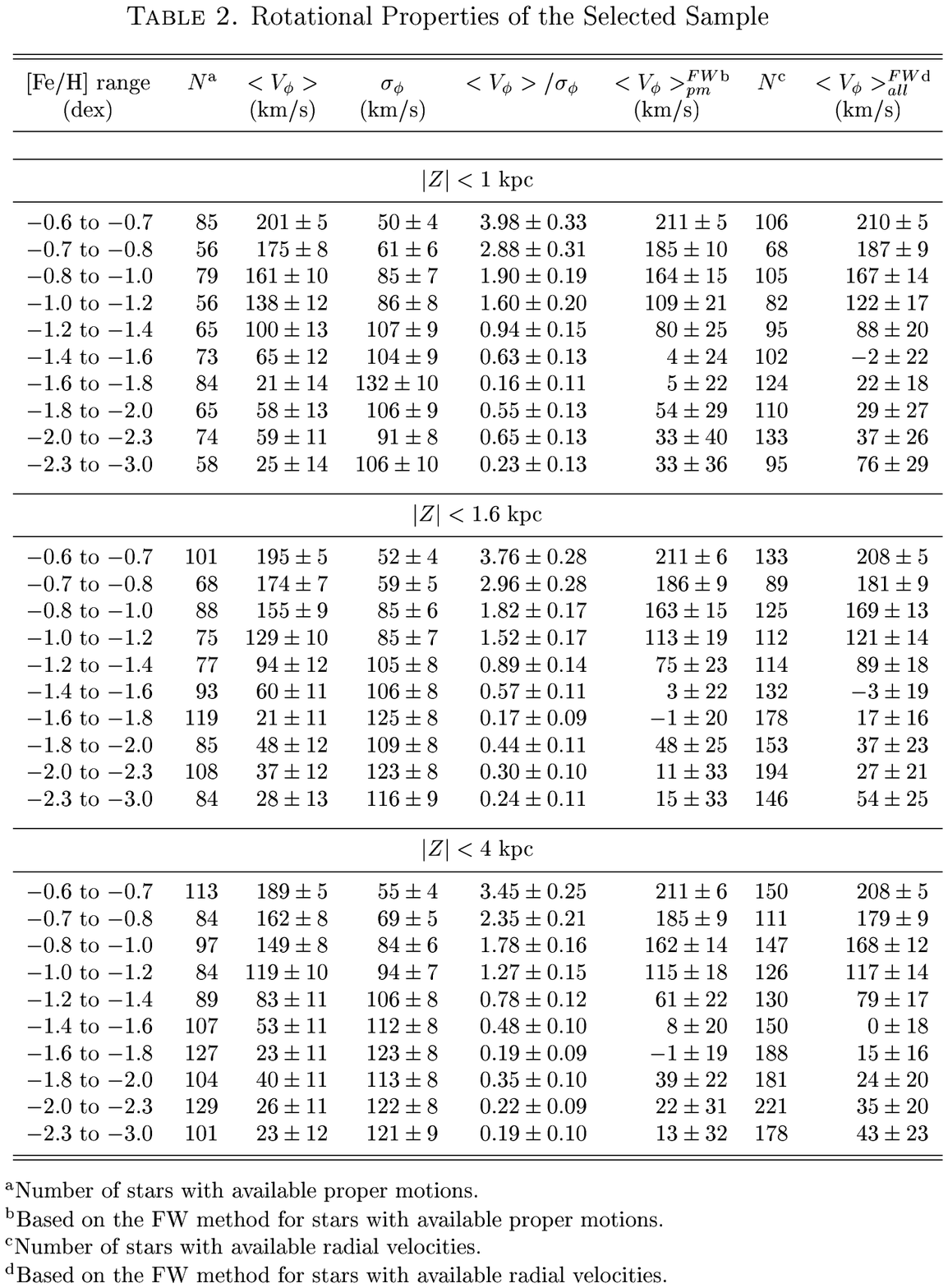}}
\centerline{\psfig{file=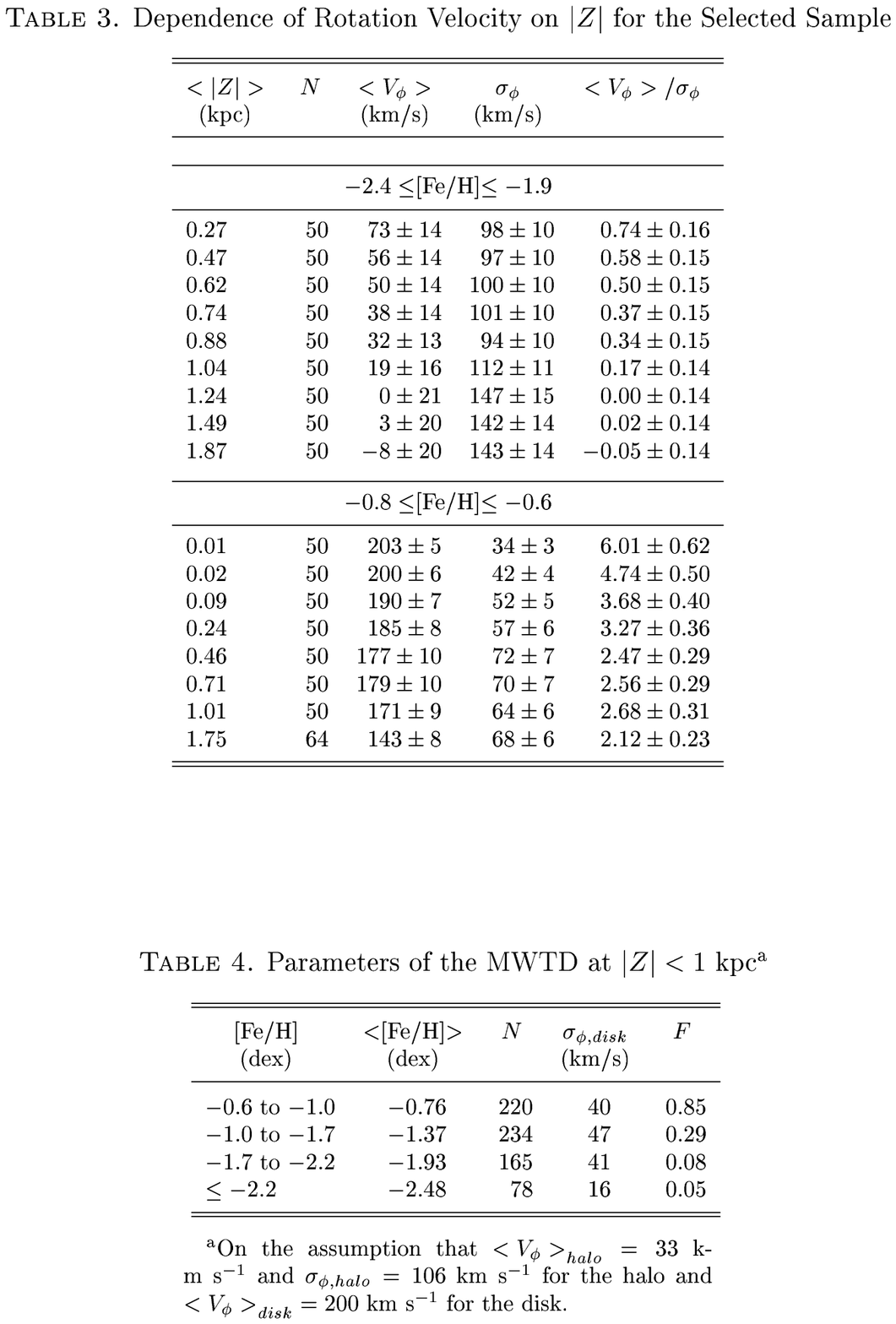}}
\centerline{\psfig{file=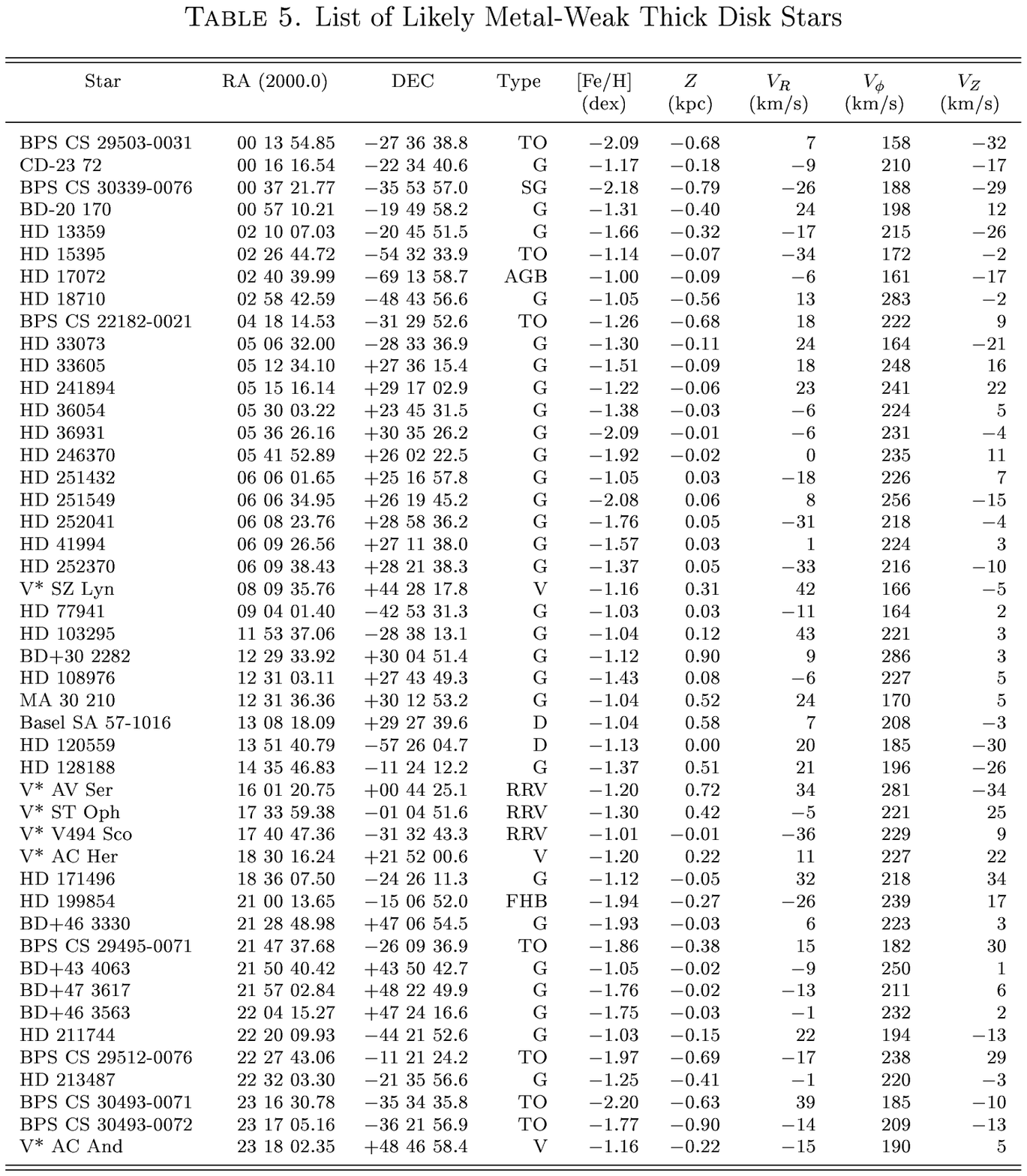}}
\centerline{\psfig{file=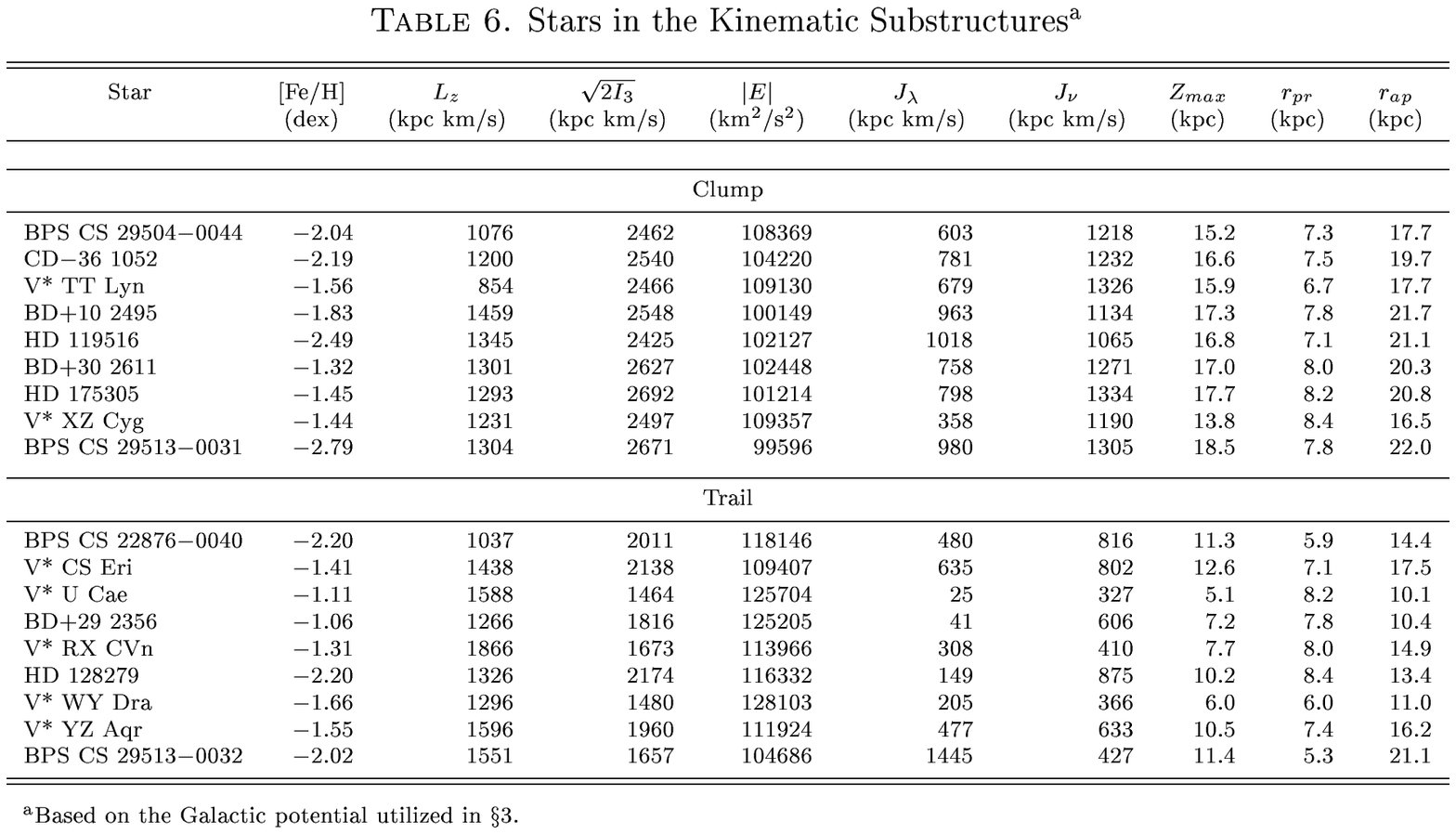}}

\clearpage
\figurenum{1}
\figcaption[]{
Distribution of the velocity components $(U,V,W)$ vs.
[Fe/H] for the 1203 stars with available proper motions. }

\figurenum{6}
\figcaption[]{
(a) The relation between [Fe/H] and $e$ for 1203 stars
with [Fe/H] $\le-0.6$.  Note the diverse range of $e$ even at low
metallicities.  (b) The relation between [Fe/H] and mean eccentricities $<e>$
(solid line with filled circles), calculated by passing a box of width $N=100$
stars (ordered by metallicity), with an overlap of 20 stars each.  The dashed
line denotes the result of Carney et al. (1996) for their high-proper-motion
sample. }

\figurenum{11}
\figcaption[]{
The relation between $V_\phi$ and $Z_{max}$ for stars
with [Fe/H] $\le-1.5$. (a) For the sample analyzed by Carney et al. (1996).
Filled circles denote their original data, whereas crosses denote our
re-calibration of the same data using the currently adopted Galactic potential.
(b) For the simulated, kinematically biased sample via a Monte Carlo method, to
reproduce the Carney et al. result.  The lower limit to the measured proper
motions is set to 0.26'' yr$^{-1}$. (c) For the present sample.  In all panels,
solid lines with error bars show local regression lines through the data.  Note
that the non-kinematically selected stars in (c) show no net rotation at large
$Z_{max}$, in contrast to the Carney et al. sample (a). }

\clearpage
\figurenum{2}
\begin{figure}
\begin{center}
\leavevmode
\epsfxsize=0.8\columnwidth\epsfbox{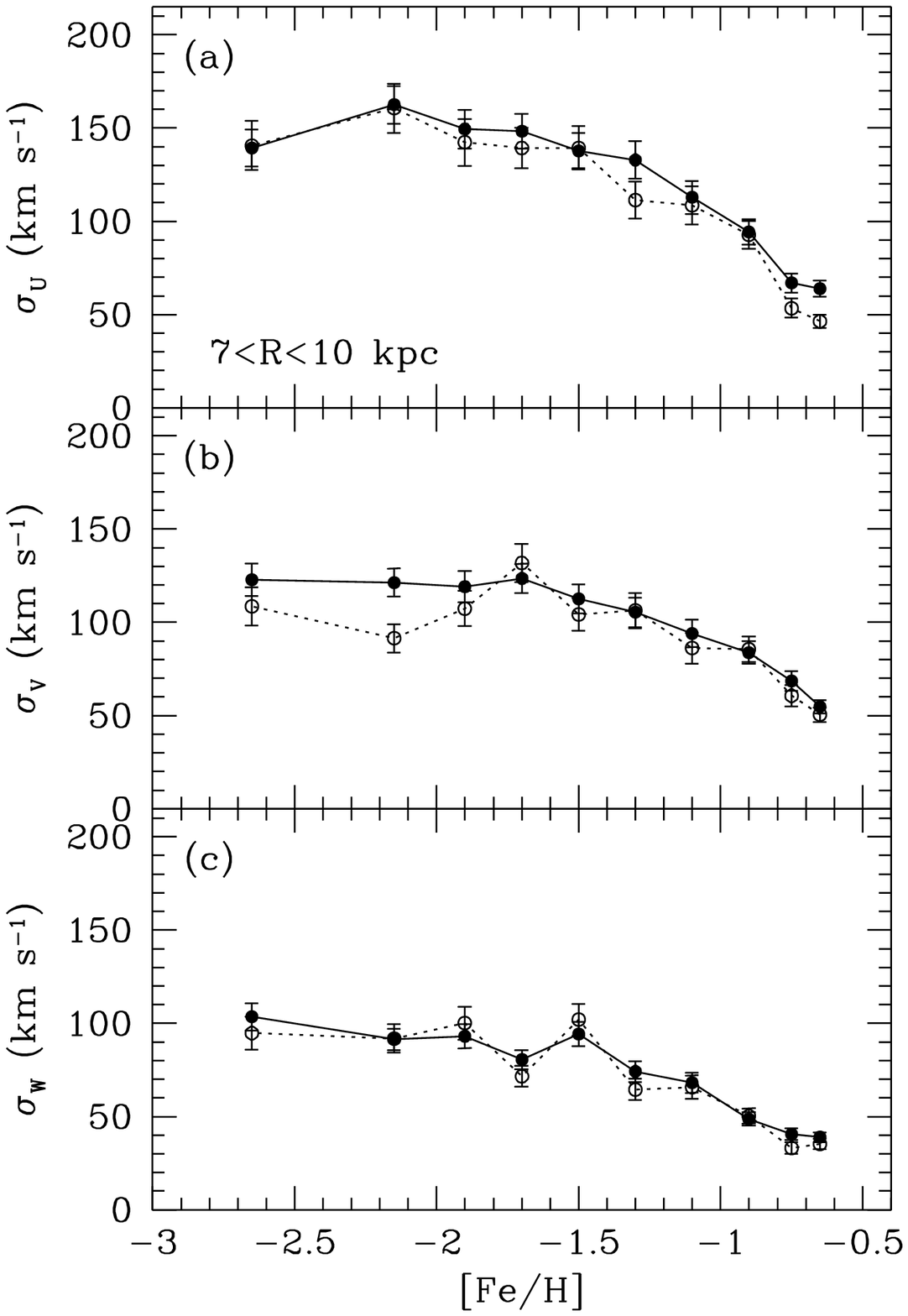}\hfil
\end{center}
\caption{
Distribution of the velocity dispersions
$(\sigma_U,\sigma_V,\sigma_W)$ vs.  [Fe/H] for the Selected Sample with
$7<R<10$ kpc, $D<4$ kpc, and $V_{RF}\le 550$ km~s$^{-1}$. Filled and open
circles denote the stars at $|Z|<1$ kpc and $|Z|<4$ kpc, respectively.
\label{fig2}}
\end{figure}

\clearpage
\figurenum{3}
\begin{figure}
\begin{center}
\leavevmode
\epsfxsize=0.8\columnwidth\epsfbox{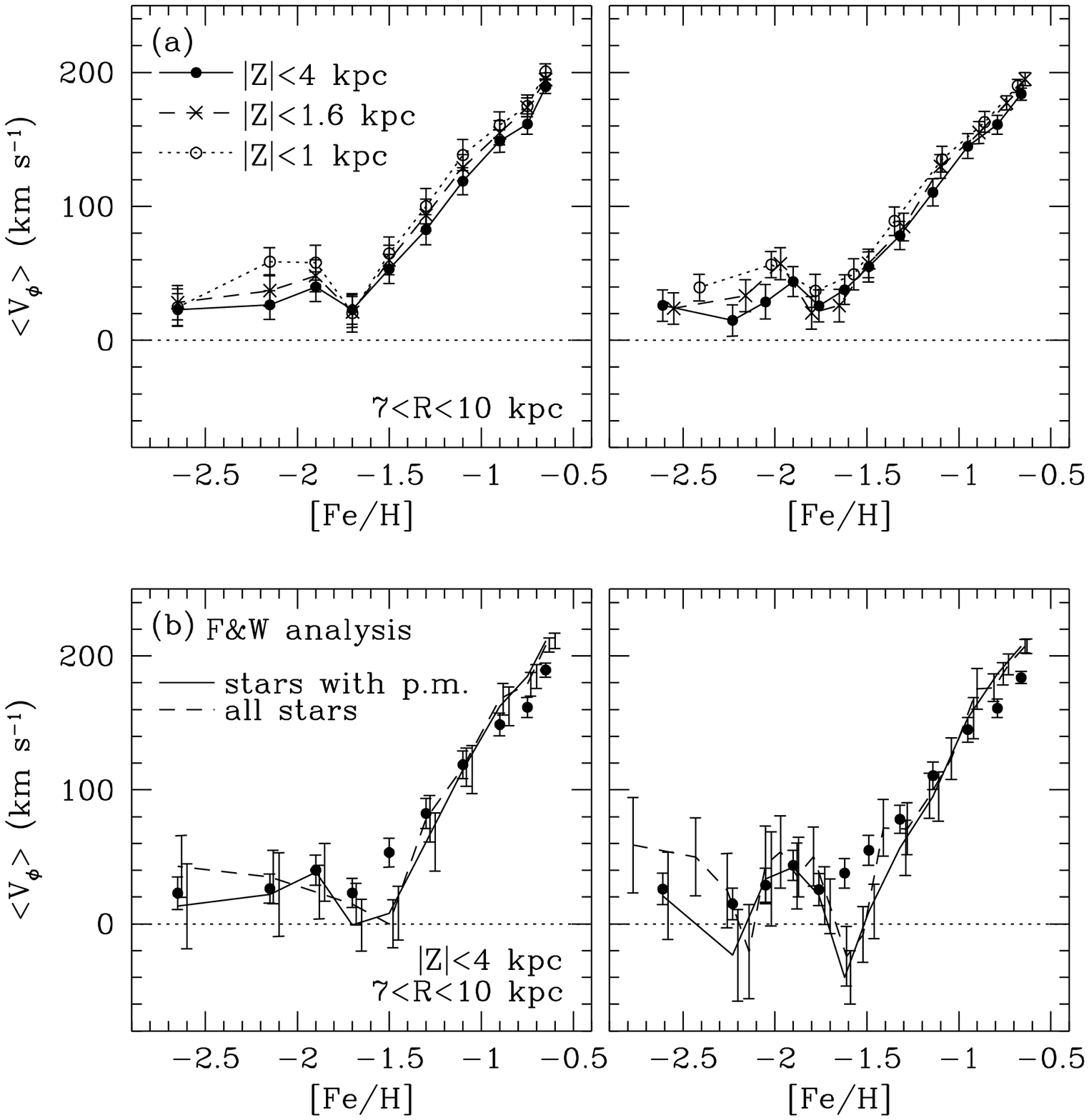}\hfil
\end{center}
\caption{
Distribution of the mean rotational velocities
$<V_\phi>$ vs. [Fe/H] for the Selected Sample, assuming an $LSR$ rotation
velocity of 220 km~s$^{-1}$. Left-hand panels show the results
for characteristic abundance ranges, whereas in the right-hand panels,
we calculate $<V_\phi>$ by passing a box of width $N=100$ stars (ordered by
metallicity), with an overlap of 20 stars each.  (a) Based on the full
knowledge of proper motions and radial velocities. Filled
circles, crosses, and open circles correspond to the stars at $|Z|<4$ kpc,
$|Z|<1.6$ kpc, and $|Z|<1$ kpc, respectively.  (b) Based on the Frenk \& White
methodology using radial velocities alone. The solid line denotes $<V_\phi>$
for the stars with available proper motions at $|Z|<4$ kpc, i.e. the same
sample as in panel (a).  The dashed line denotes $<V_\phi>$ for the stars with
available radial velocities and $|Z|<4$ kpc. For comparison, we plot the filled
circles from panel (a). \label{fig3}}
\end{figure}

\clearpage
\figurenum{4}
\begin{figure}
\begin{center}
\leavevmode
\epsfxsize=0.8\columnwidth\epsfbox{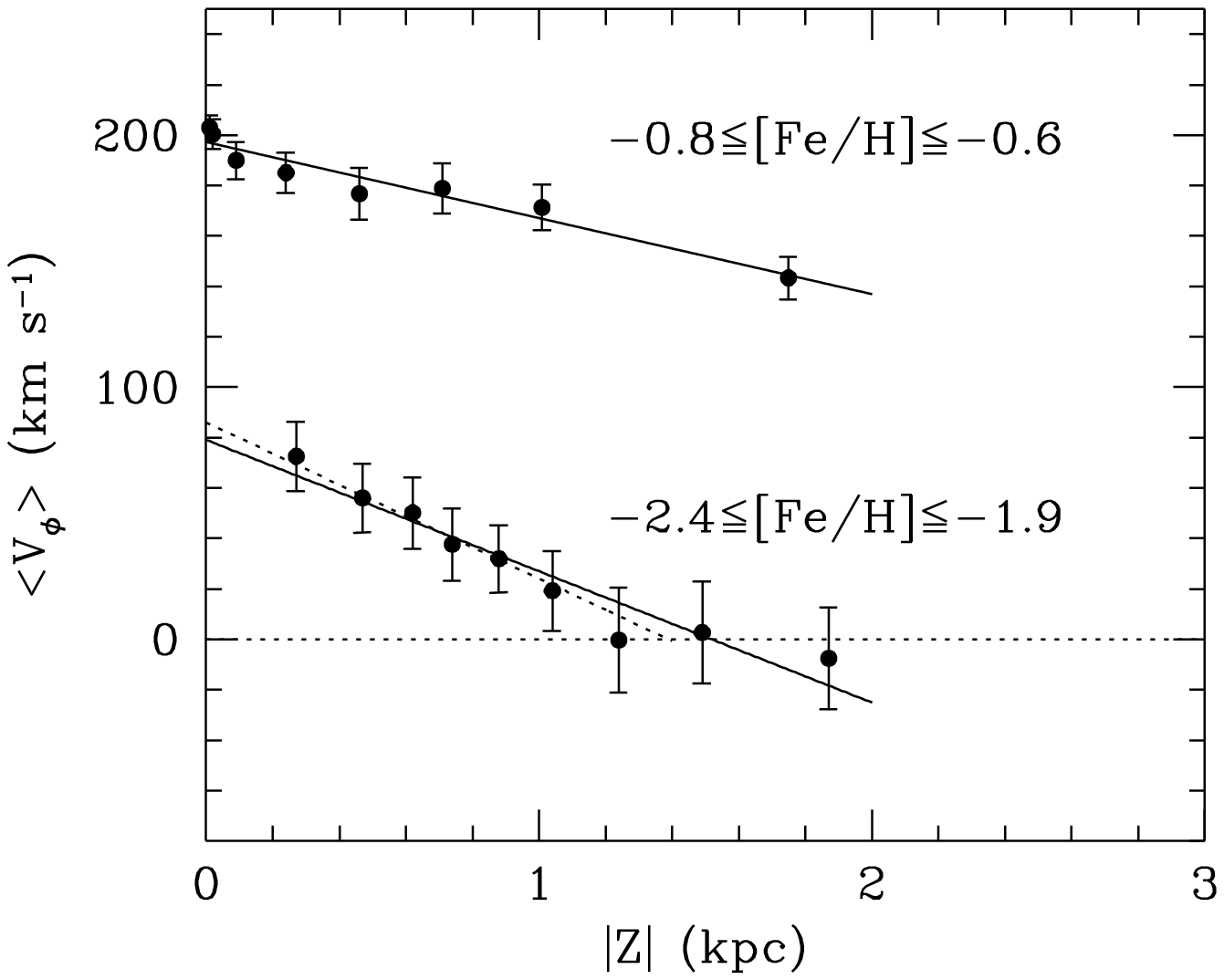}\hfil
\end{center}
\caption{
Distribution of the mean rotational velocities
$<V_\phi>$ vs. $|Z|$ for stars of the Selected Sample, in the abundance ranges
$-2.4\le$ [Fe/H] $\le-1.9$ and $-0.8\le$ [Fe/H] $\le-0.6$, respectively. The
binning in $|Z|$ is made by sweeping a box of 50 stars through the data
(ordered by $|Z|$), with an overlap of 30 stars. Solid lines are least-square
fits to the data.  The dotted line for $-2.4\le$ [Fe/H] $\le -1.9$ is the fit
obtained after excluding the last point at $|Z|=1.76$ kpc. \label{fig4}}
\end{figure}

\clearpage
\figurenum{5}
\begin{figure}
\begin{center}
\leavevmode
\epsfxsize=0.8\columnwidth\epsfbox{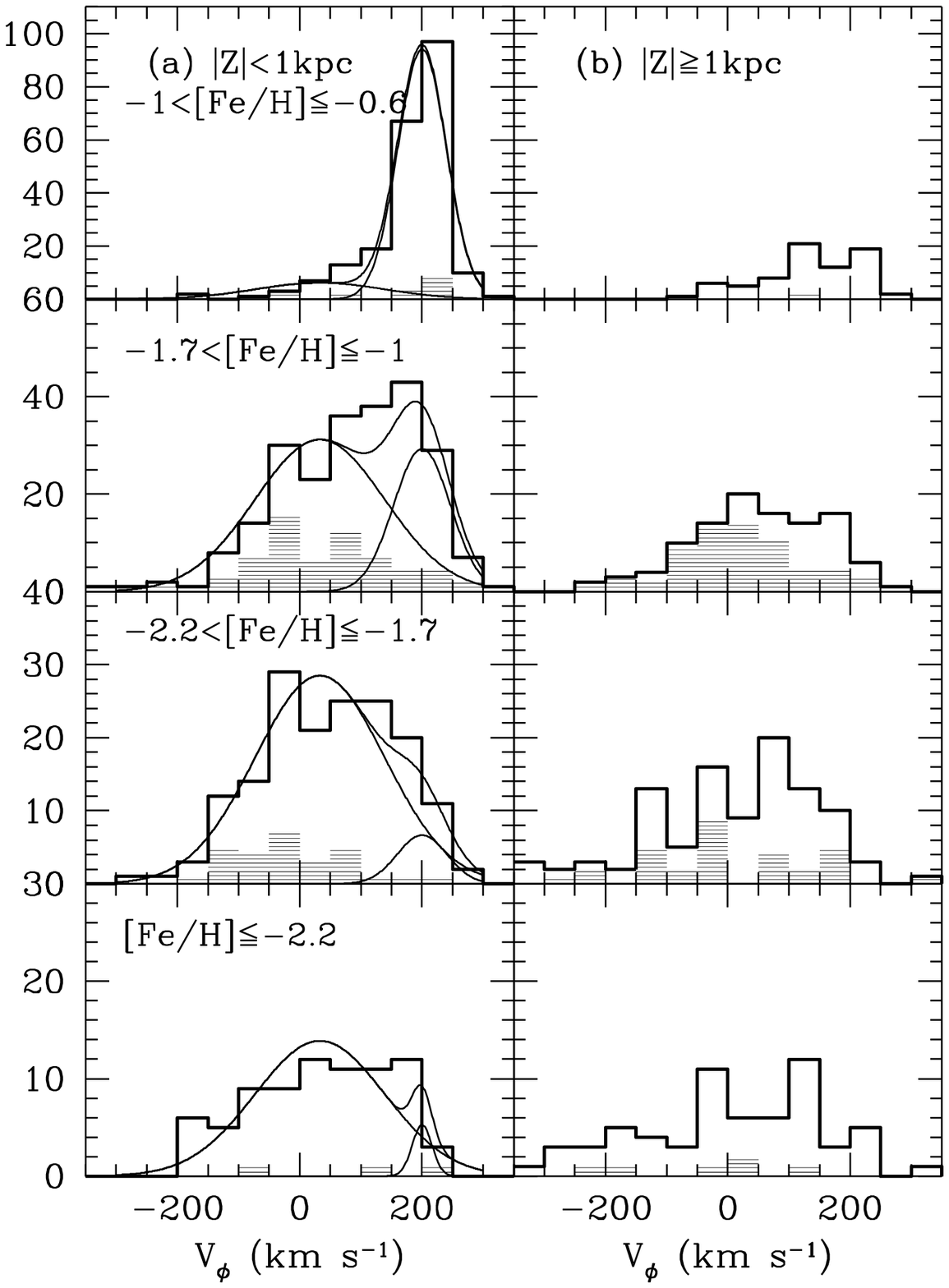}\hfil
\end{center}
\caption{
Distribution of $V_\phi$ in different metallicity ranges for the stars with
available proper motions. Solid and shaded histograms denote all stars and RR
Lyrae stars, respectively, for (a) $|Z|<1$ kpc  and (b) $|Z|\ge1$ kpc. Also
shown in the left-hand panels are the results of the maximum likelihood
analysis for reproducing the $V_\phi$ distribution at $|Z|<1$ kpc, based on a
mixture of two Gaussian components, which we associate with the halo and thick
disk. \label{fig5}}
\end{figure}

\clearpage
\figurenum{7}
\begin{figure}
\begin{center}
\leavevmode
\epsfxsize=0.7\columnwidth\epsfbox{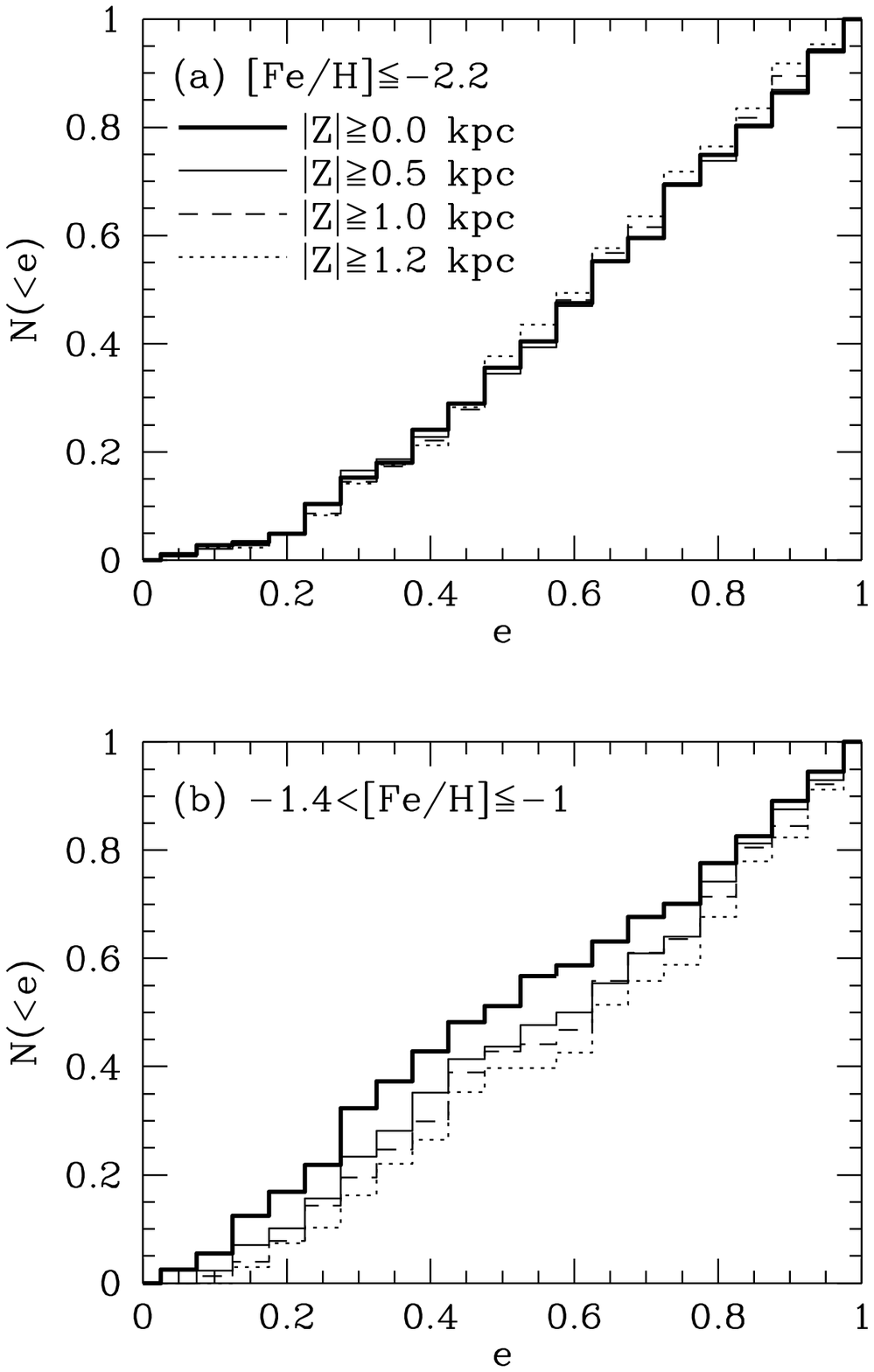}\hfil
\end{center}
\caption{
Cumulative $e$ distributions, $N(<e)$, in the two abundance ranges (a) [Fe/H]
$\le-2.2$, and (b) $-1.4<$[Fe/H] $\le-1$. The thick solid, thin solid, dashed,
and dotted histograms denote the stars at $|Z|\ge0.0$ kpc (all stars),
$|Z|\ge0.5$ kpc, $|Z|\ge1.0$ kpc, and $|Z|\ge1.2$ kpc, respectively.
\label{fig7}}
\end{figure}

\clearpage
\figurenum{8}
\begin{figure}
\begin{center}
\leavevmode
\epsfxsize=0.8\columnwidth\epsfbox{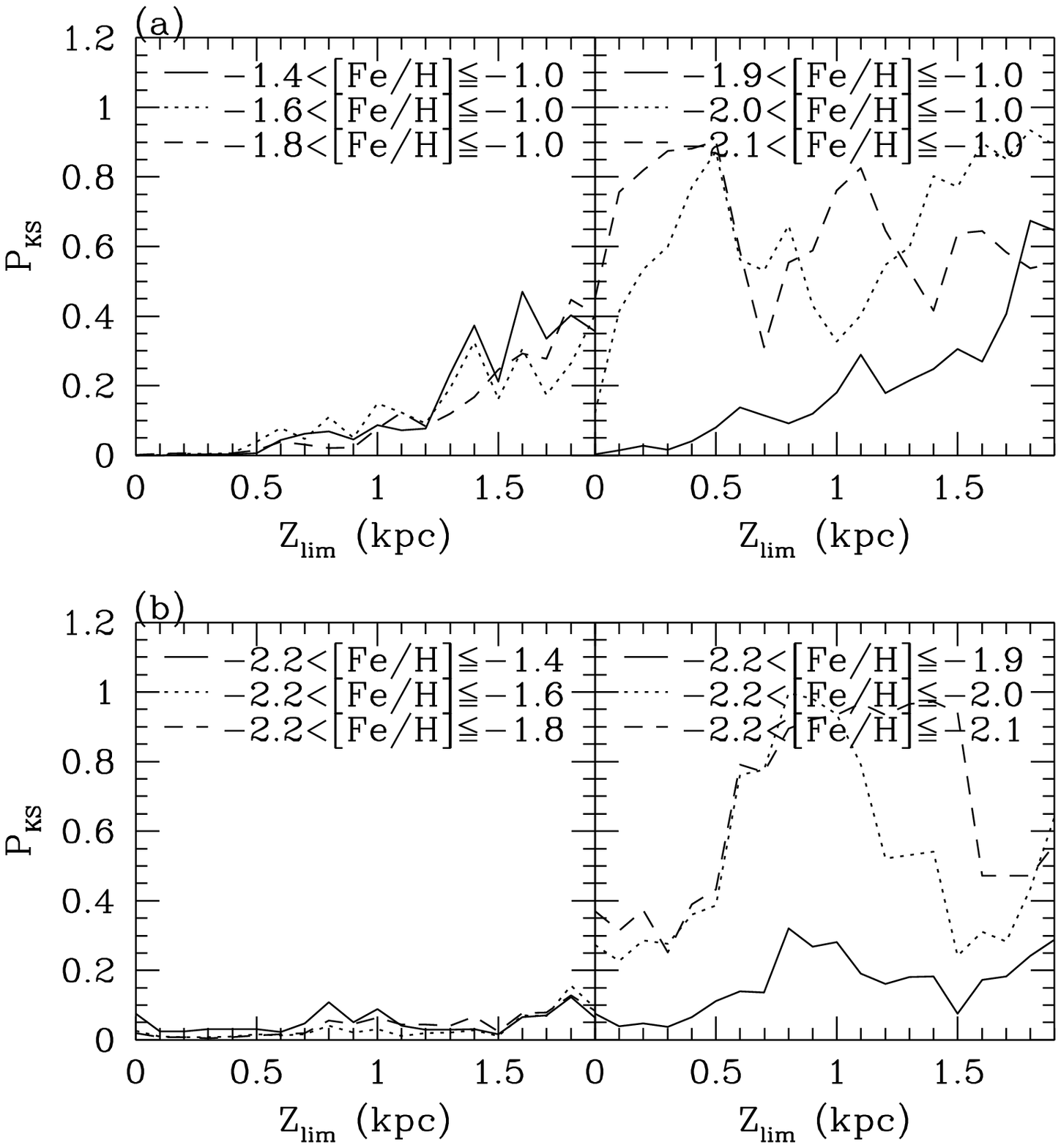}\hfil
\end{center}
\caption{
(a) Kolmogorov-Smirnoff probabilities for the stars
included in each abundance range at $|Z|>Z_{lim}$, for evaluation of the null
hypothesis that the differential $e$ distribution, $n(e)$, is drawn from the
same parent population as the ``pure'' halo with [Fe/H] $\le-2.2$ and $|Z|<1$
kpc.  The lines denote results obtained when the lower limit of each abundance
range is varied, while the upper limit is fixed at [Fe/H] $=-1$.  (b) Same as
in (a), but for the more metal-poor range.  The lines denote results obtained
when the upper limit of each abundance range is varied, while the lower limit
is fixed at [Fe/H] $=-2.2$. \label{fig8}}
\end{figure}

\clearpage
\figurenum{9}
\begin{figure}
\begin{center}
\leavevmode
\epsfxsize=0.8\columnwidth\epsfbox{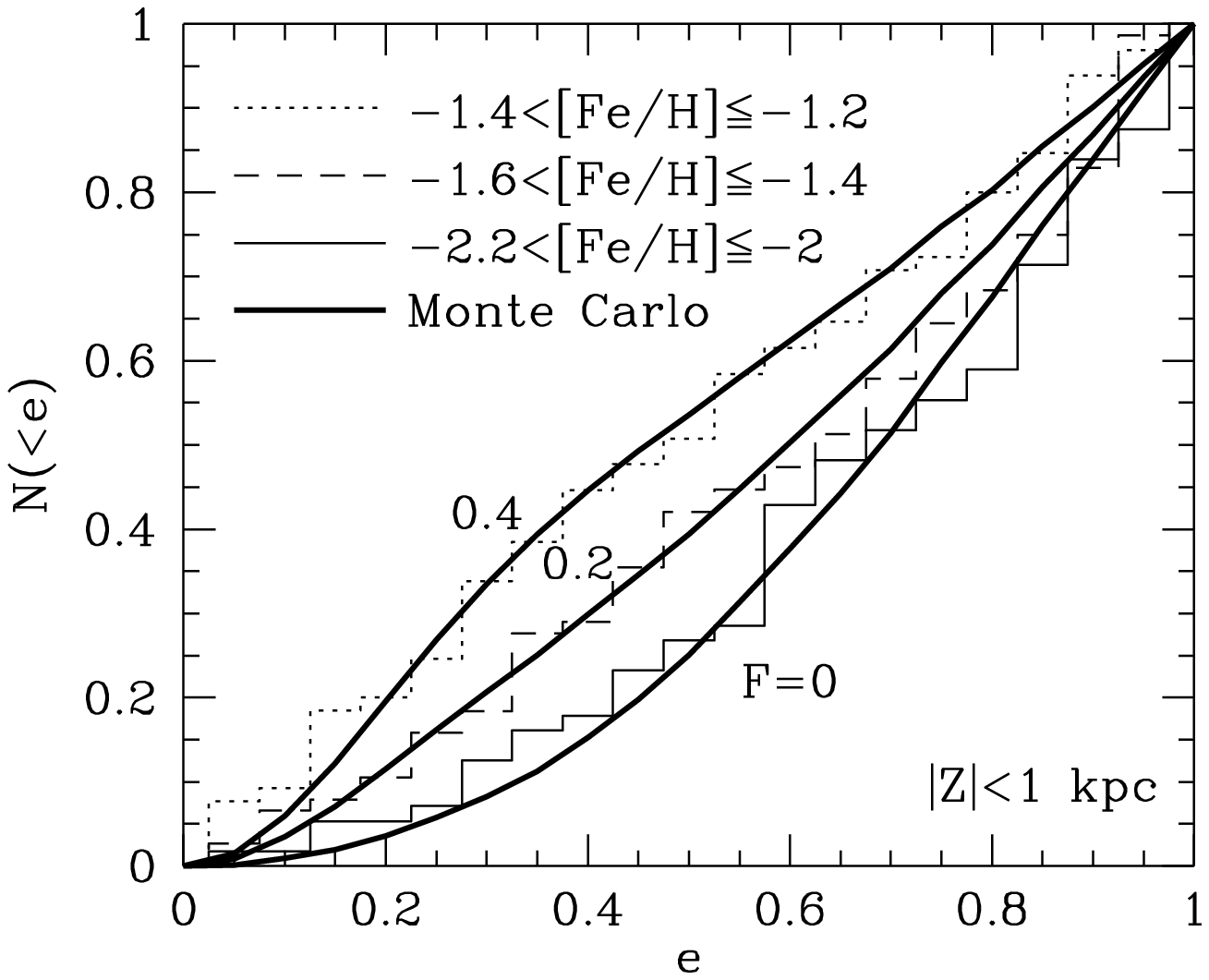}\hfil
\end{center}
\caption{
Comparisons of observed $N(<e)$ in the solar
neighborhood in various abundance ranges with Monte Carlo results, based on a
mixture of two Gaussian components, taken to represent the halo and thick disk,
where the disk fraction is denoted as $F$.  We take $<V_\phi>= +33$ km~s$^{-1}$
and $(\sigma_U,\sigma_V,\sigma_W)= (141,106,94)$ km~s$^{-1}$ for the halo, and
$<V_\phi>= +200$ km~s$^{-1}$ and $(\sigma_U,\sigma_V,\sigma_W)= (46,50,35)$
km~s$^{-1}$ for the thick disk. \label{fig9}}
\end{figure}

\clearpage
\figurenum{10}
\begin{figure}
\begin{center}
\leavevmode
\epsfxsize=0.8\columnwidth\epsfbox{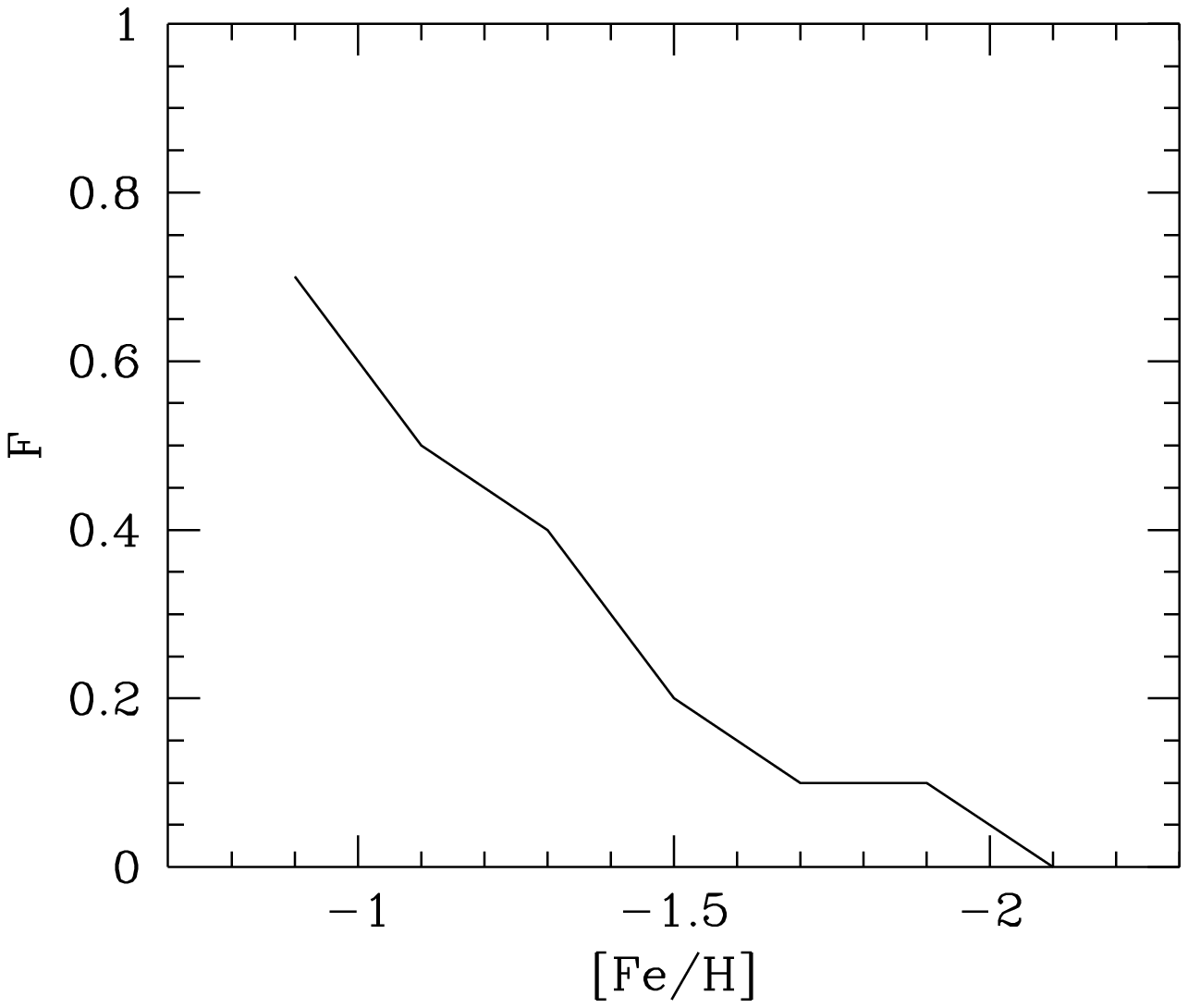}\hfil
\end{center}
\caption{
The fraction $F$ of the thick disk at $|Z|<1$ kpc as a
function of [Fe/H], derived by the comparison between the observed and
predicted $N(<e)$. \label{fig10}}
\end{figure}

\clearpage
\figurenum{12}
\begin{figure}
\begin{center}
\leavevmode
\epsfxsize=0.7\columnwidth\epsfbox{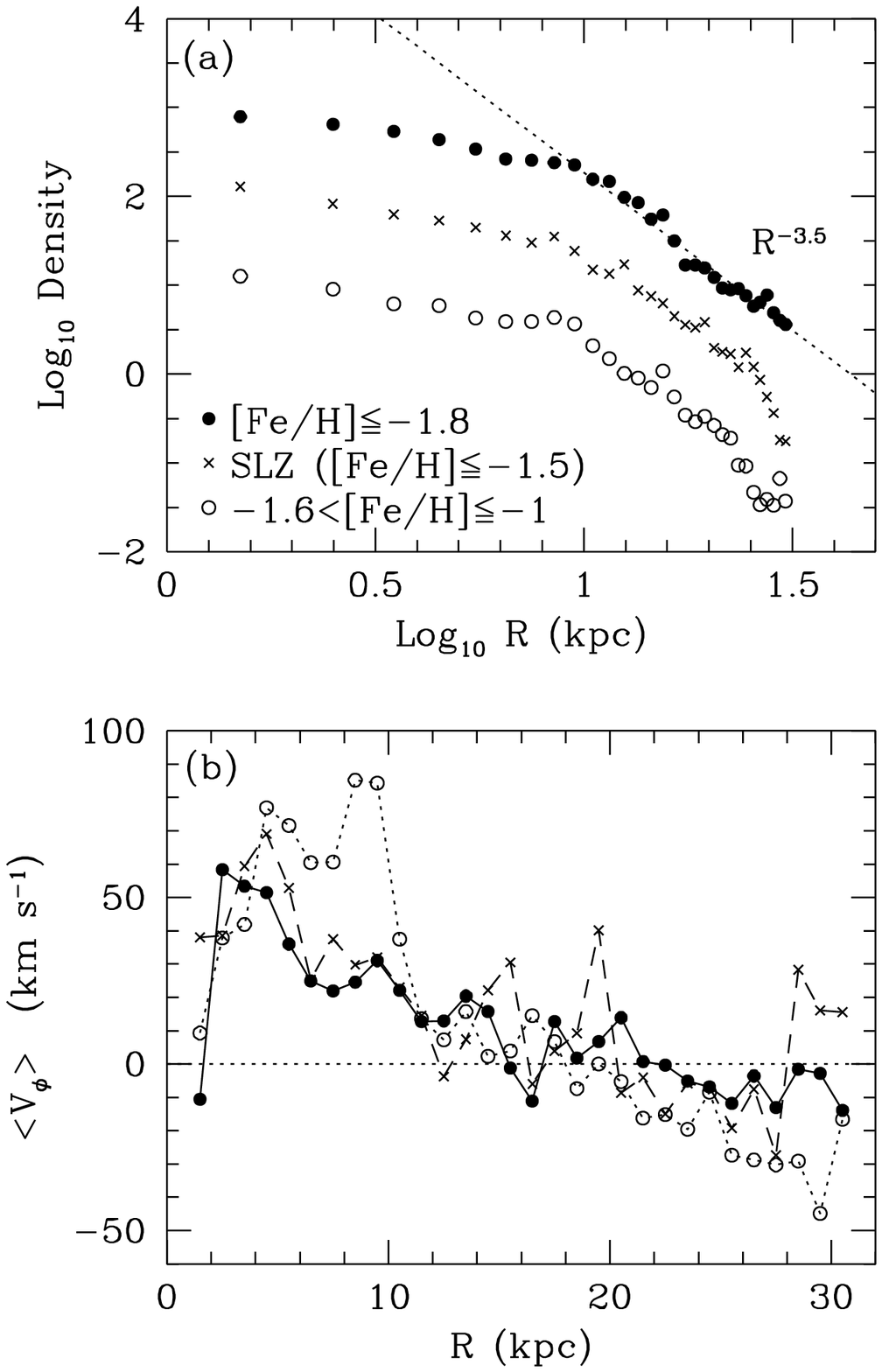}\hfil
\end{center}
\caption{
(a) Density distributions of the reconstructed halo in
the Galactic plane, for [Fe/H] $\le-1.8$ (filled circles) and $-1.6<$ [Fe/H]
$\le-1$ (open circles).  Both plots have been shifted arbitrarily along the
vertical axis.  The dotted line denotes the power-law model with exponent
$\beta=-3.5$.  For comparison, the density distribution based on the SLZ sample
with [Fe/H] $\le-1.5$ is shown as crosses.  (b) Mean rotational velocities of
the reconstructed halo projected onto the Galactic plane.  Symbols are the same
as in (a). \label{fig12}}
\end{figure}

\clearpage
\figurenum{13}
\begin{figure}
\begin{center}
\leavevmode
\epsfxsize=0.5\columnwidth\epsfbox{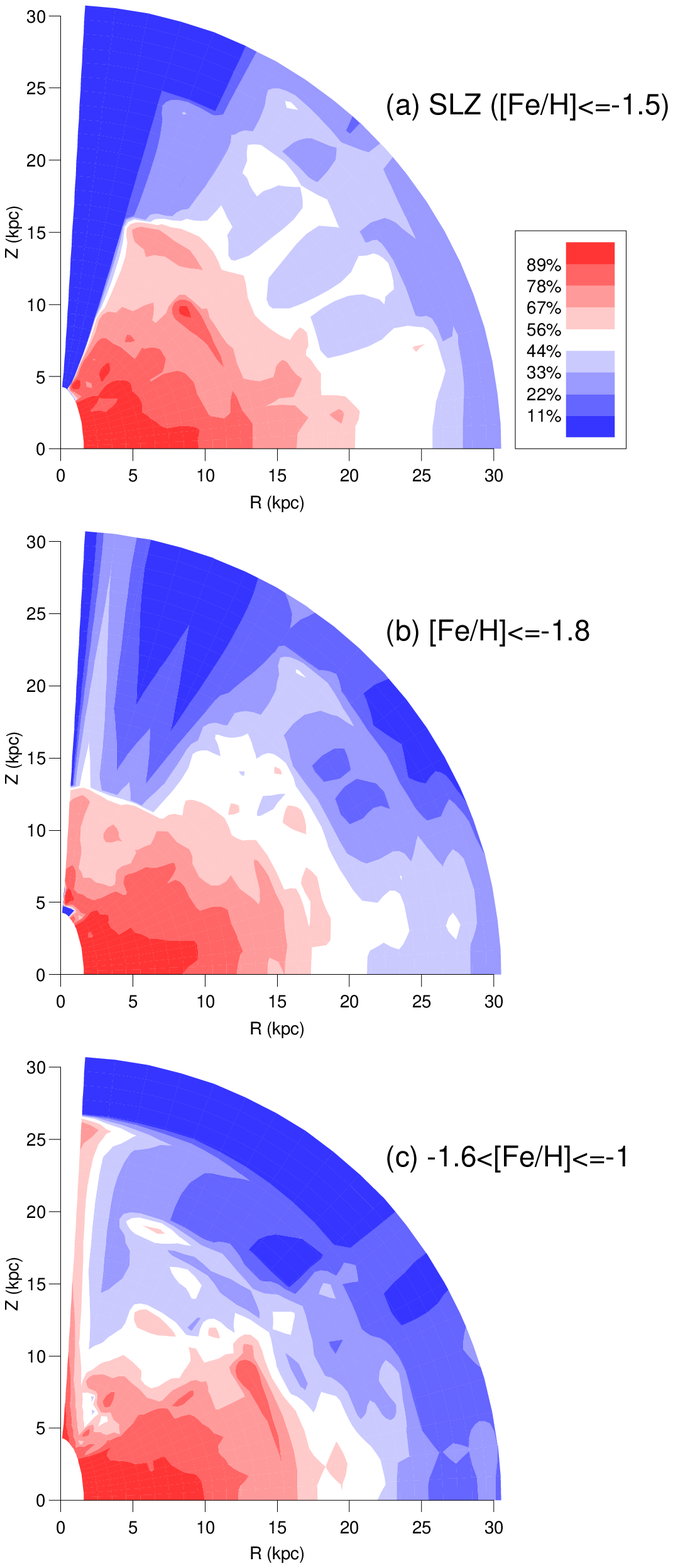}\hfil
\end{center}
\caption{
Equidensity contours for the reconstructed halo in
the $(R,Z)$ plane. (a) For the SLZ sample with [Fe/H] $\le-1.5$, (b) for 
our sample with [Fe/H] $\le-1.8$, and (c) for our sample with $-1.6<$ [Fe/H]
$\le-1$.
\label{fig13}}
\end{figure}

\clearpage
\figurenum{14}
\begin{figure}
\begin{center}
\leavevmode
\epsfxsize=0.6\columnwidth\epsfbox{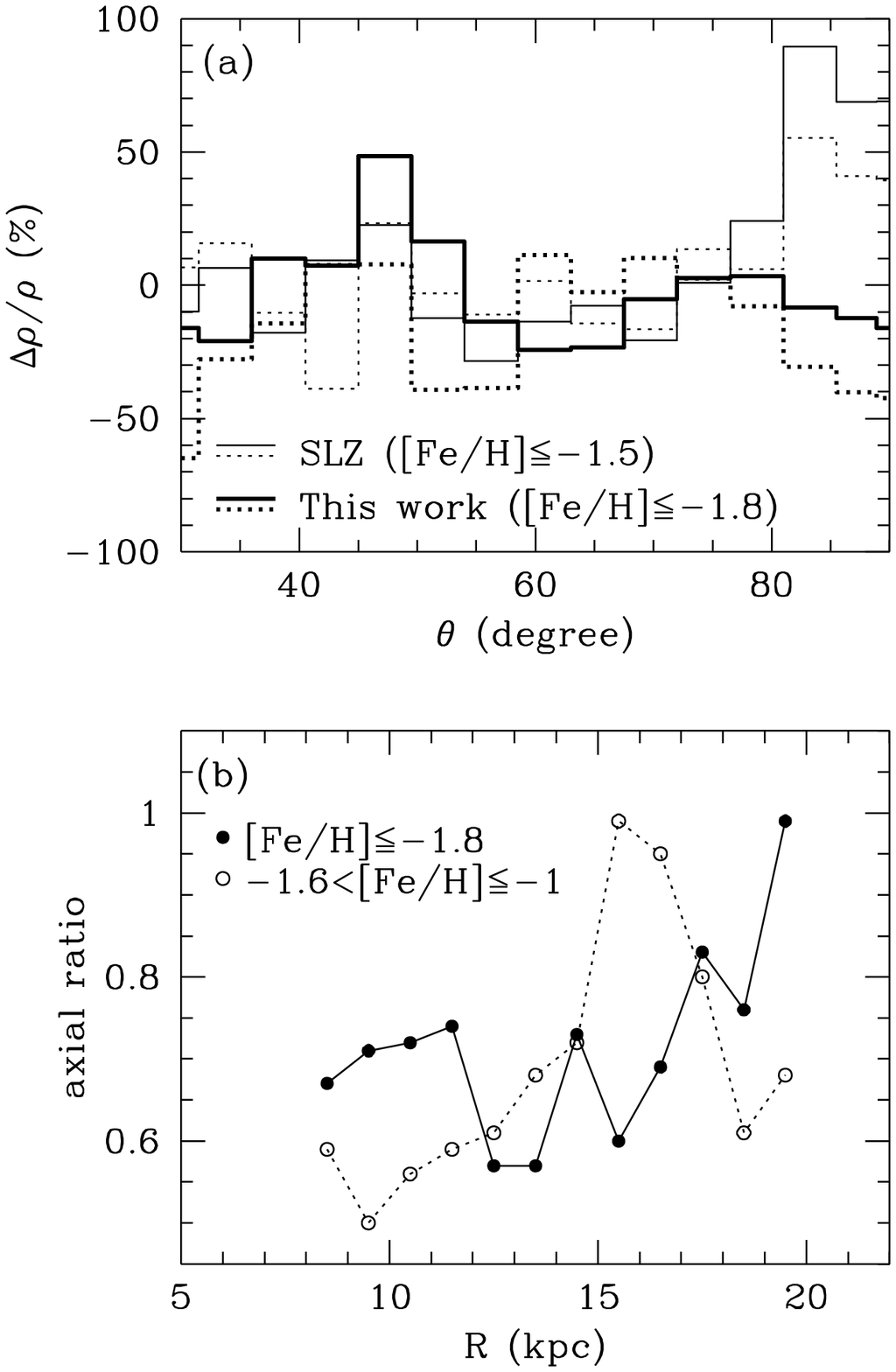}\hfil
\end{center}
\caption{
(a) Relative density deviations along elliptical fits
to equidensity contours with $a=10.5$ kpc (solid lines) and $a=13.5$ kpc
(dotted lines), when the fits are made by omitting the data points near the
plane ($\theta>80^\circ$). The thin lines denote the SLZ sample with [Fe/H]
$\le-1.5$ ($q=0.85$ for $a=10.5$ kpc and $q=0.82$ for $a=13.5$ kpc), whereas
the thick lines denote our sample with [Fe/H] $\le-1.8$ ($q=0.70$ for $a=10.5$
kpc and $q=0.51$ for $a=13.5$ kpc). Note that contrary to SLZ, our sample does
not show a density excess near the plane.  (b) Axial ratios for the density
distribution of the reconstructed halo, based on the elliptical fits to
equidensity contours including stars near the plane. The filled and open
circles denote our sample with [Fe/H] $\le-1.8$ and $-1.6<$ [Fe/H] $\le-1$,
respectively. \label{fig14}}
\end{figure}

\clearpage
\figurenum{15}
\begin{figure}
\begin{center}
\leavevmode
\epsfxsize=0.8\columnwidth\epsfbox{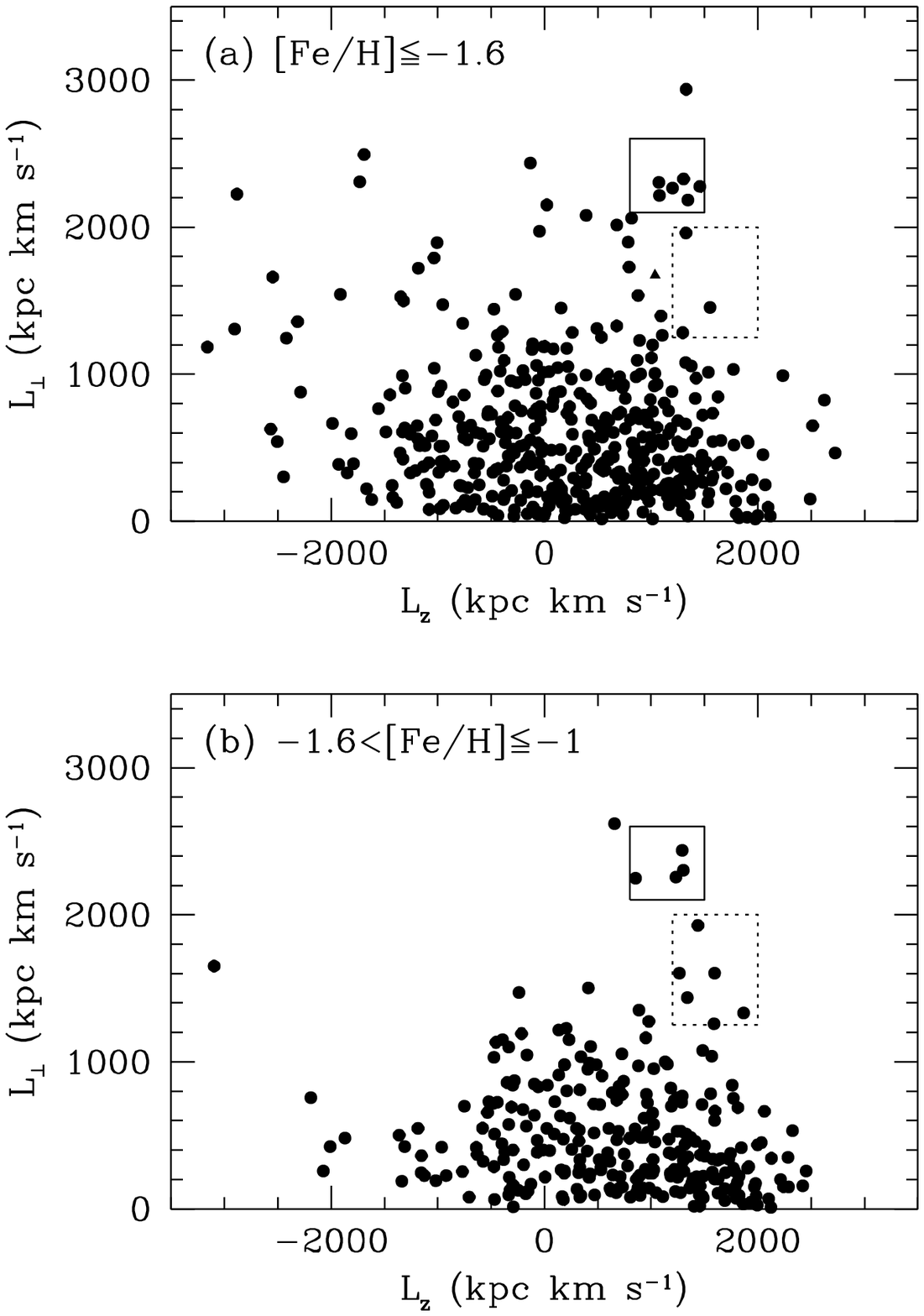}\hfil
\end{center}
\caption{
Distribution of the stars with $D<2.5$ kpc in the
angular momentum components $L_z$ vs. $L_\perp = (L_x^2+L_y^2)^{1/2}$, for
(a) [Fe/H] $\le-1.6$ and (b) $-1.6<$ [Fe/H] $\le-1$. The solid and dotted boxes
denote the regions of the ``clump'' and ``trail'' as defined in the text.  A
triangle in panel (a) denotes BPS CS 22876-0040, which is assigned to a
``trail'' member, as discussed in the text.
\label{fig15}}
\end{figure}

\clearpage
\figurenum{16}
\begin{figure}
\begin{center}
\leavevmode
\epsfxsize=0.8\columnwidth\epsfbox{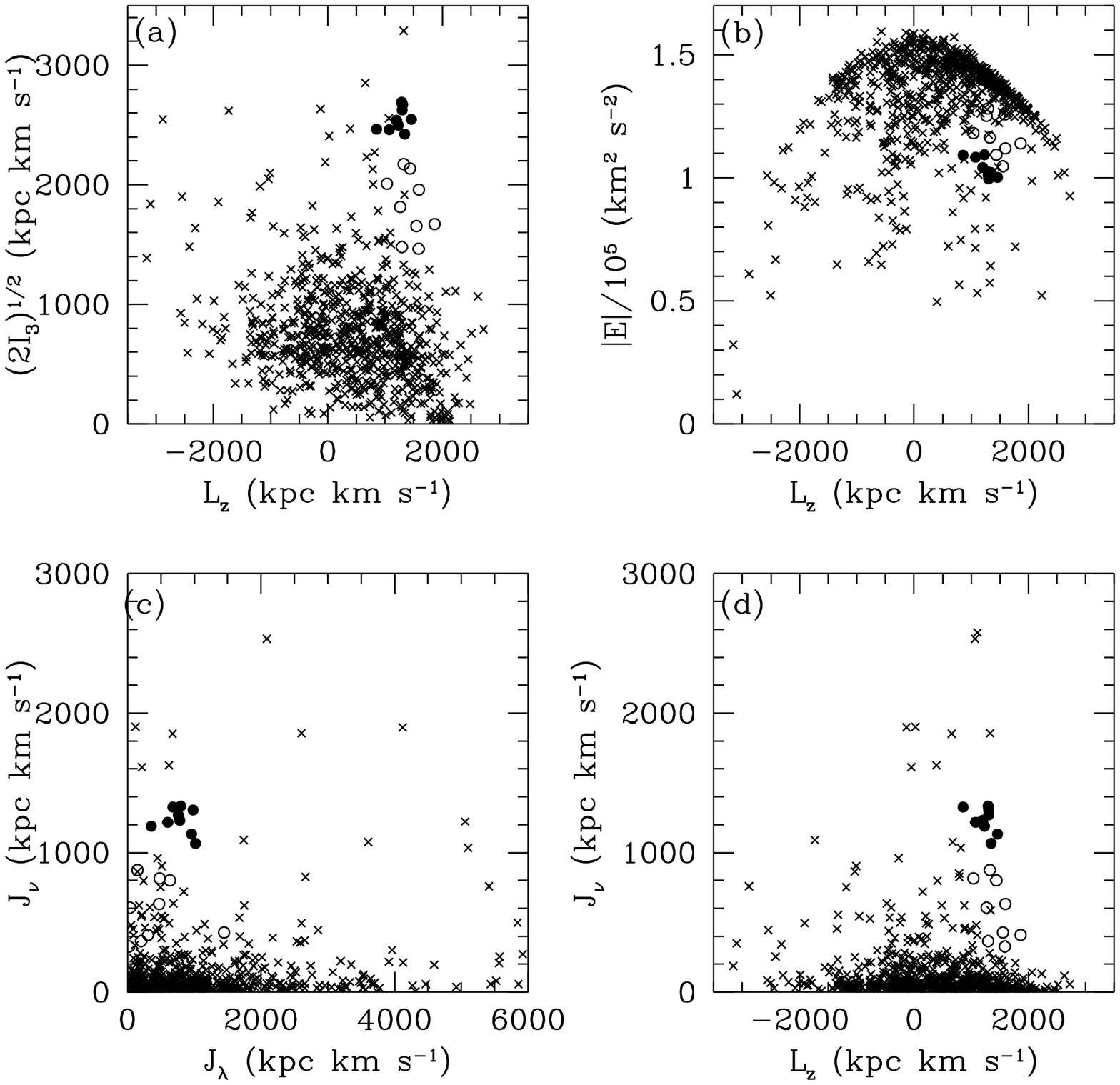}\hfil
\end{center}
\caption{
Distributions of the stars with $D<2.5$ kpc and [Fe/H]
$\le-1$, in the planes of (a) $(2I_3)^{1/2}$ vs. $L_z$, (b) $|E|$ vs.  $L_z$,
(c) $J_\nu$ vs. $J_\lambda$, and (d) $J_\nu$ vs. $L_z$.  The filled and open
circles denote the stars in the ``clump'' and ``trail'' regions, whereas
crosses denote the rest of the stars.  Note that we exclude HD~214161 from the
``clump'' and CS~Ser from the ``trail'', because these stars are located at
quite different regions from other member stars in the $J_\nu$ vs. $J_\lambda$
diagram. \label{fig16}}
\end{figure}

\clearpage
\figurenum{17}
\begin{figure}
\begin{center}
\leavevmode
\epsfxsize=0.7\columnwidth\epsfbox{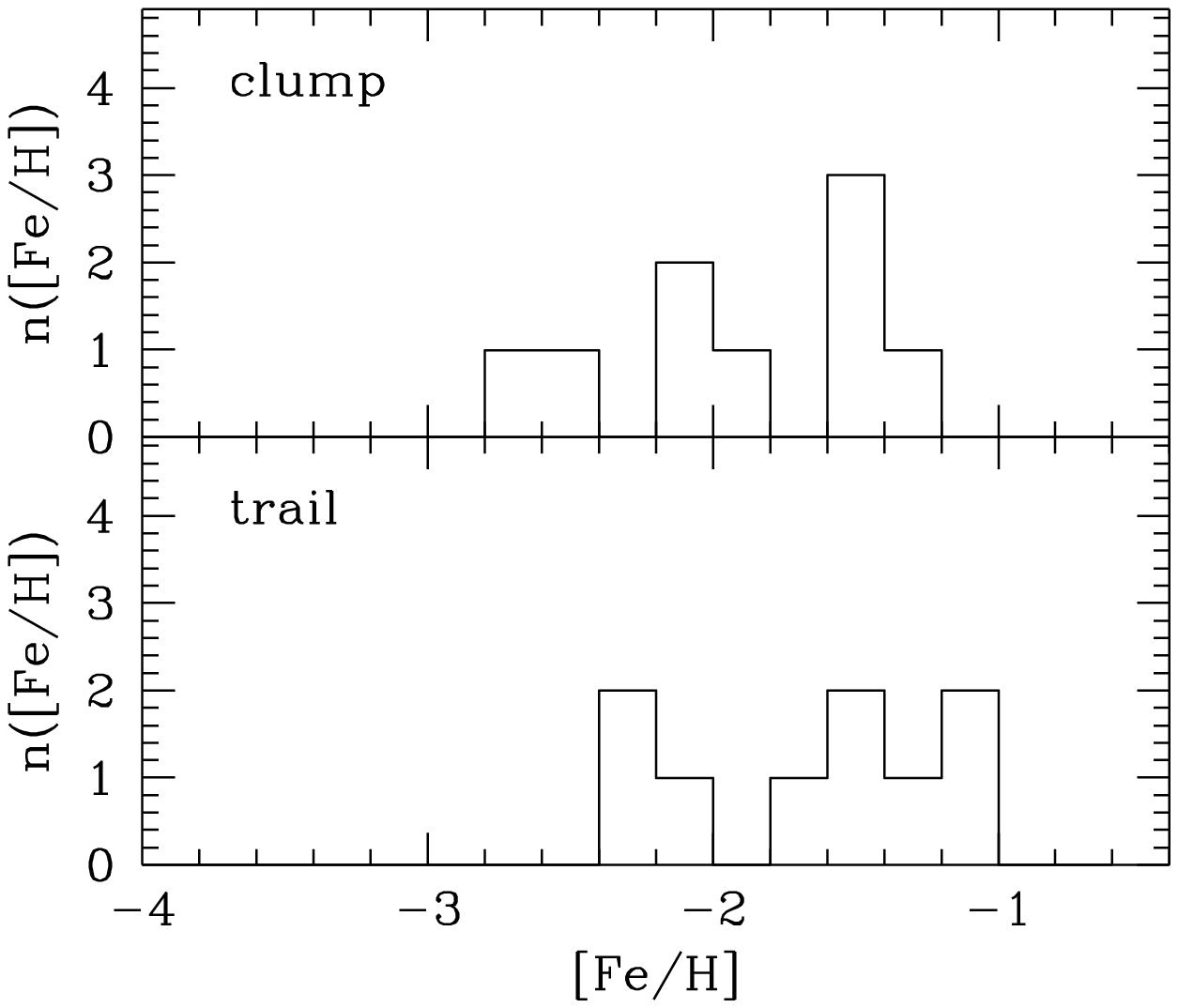}\hfil
\end{center}
\caption{
Metallicity distributions of the stars in the ``clump''
and ``trail'' regions. \label{fig17}}
\end{figure}

\end{document}